\documentclass{aa501}     
\usepackage{graphicx}
\def\cm2{\,{\rm cm^{-2}}}
\def\kms{\,{\rm {km\,s^{-1}}}}

\def\Msun{\rm M_{\odot}}
\def\Lsun{\rm L_{\odot}}

\def\aua{{\rm A\&A} }
\def\auas{{\rm A\&AS} }
\def\apj{{\rm ApJ} }
\def\aj{{\rm AJ} }
\def\apjs{{\rm ApJS} }

\def\araa{{\rm ARAA} }
\def\mnras{{\rm MNRAS} }

\begin{document}
 
\title{Neutral hydrogen in dwarf galaxies}
 
   \subtitle{I. The spatial distribution of HI}
 
\author{J.M. Stil
          \inst{1, 2}
           and F. P. Israel
          \inst{1}
           }
 
   \offprints{F.P. Israel}
 
  \institute{Sterrewacht Leiden, P.O. Box 9513, 2300 RA Leiden,
             The Netherlands
  \and       Physics Department, Queen's University, Kingston ON K7L 4P1, 
	     Canada}
 
\date{received ????; accepted ????}
 
\abstract{This paper is the first in a series presenting a sample of 
30 late-type dwarf galaxies, observed with the Westerbork Synthesis 
Radio Telescope (WSRT) in the 21-cm line of neutral atomic hydrogen (HI). 
The sample itself, the HI content of and the HI distribution in the sample 
galaxies are briefly discussed.  Four sample galaxies were also
detected in the continuum.
}

\maketitle
 
\section{Introduction.}        

Galaxies come in a wide variety of shapes and sizes. The larger
galaxies are usually accompanied by a number of smaller (dwarf)
galaxies, although dwarf galaxies also occur by themselves.
Late-type dwarf galaxies are generally rich in neutral atomic hydrogen 
(HI) gas, usually more so than much larger late type spiral
galaxies. Their optical luminosity can vary considerable. Blue,
compact dwarf galaxies (BCGD) which appear subject to intense star formation
are relatively easy to observe. More quiescent, redder galaxies
are not so easy to find, especially if they have low surface 
brightnesses (LSB). The ratio of HI-mass to light is higher in dwarf 
galaxies than in much larger galaxies of high luminosity (e.g. Roberts 
$\&$ Haynes 1994). The 21-cm HI line is therefore an excellent tool 
for finding dwarf galaxies many of which otherwise might escape attention. 
Many HI line surveys exist, for example those by Fisher $\&$ Tully
(1975, 1981), Thuan $\&$ Seitzer (1979), Thuan $\&$ Martin (1981),
Hoffman et al. (1987). However, only radio interferometers provide 
sufficient spatial resolution to study the actual HI structure of 
dwarf galaxies. This is important, because relatively easily
observed neutral hydrogen is the major constituent of the interstellar
medium in galaxies (cf. Israel, 1988). 

\section{The galaxy sample}

In constructing our observing sample, we have based ourselves on the
compilation by Melisse $\&$ Israel (1994) of galaxies classified Im and
Sm, and limited ourselves to those galaxies that are in the northern 
hemisphere , i.e. have declinations above 14$^{\circ}$ so as to be 
observable with the WSRT). There is no single unambiguous definition of 
a dwarf galaxy. Often, a galaxy is considered to be a dwarf if its 
absolute luminosity corresponds to the light of no more than half a 
billion suns ($M_{\rm B} > -16$), about one per cent of the luminosity 
of a spiral galaxy such as the Milky Way or M~31. Although guided by 
this definition, we have not strictly adhered to it. Rather, we have
selected a sample of mostly Im galaxies primarily chosen to cover a 
range of optical properties, in particular colour. Sample galaxy 
colours range from $(B-V) = 0$ to $(B-V) = 0.7$. No attempt was made 
to define a complete sample. Basic 
information on the sample galaxies is given in Table~\ref{phys-sample-tab}.

%
\begin{table*}
\caption[]{Basic data on sample dwarf galaxies}
\begin{center}
\begin{tabular}{|  l  | l  l  | c c | l c r | l | l |} 
\hline 
Name  & \multicolumn{2}{c}{Fringe Stopping Center} \vline &  dist. & $\rm Q_{dist}$  &\ \  $M_B$ &{\footnotesize $B-V$}& log$FIR$ & \ \ \ \ \ group & Other names \\
\hline 
      & $\alpha(1950)$  &  \ \ \ $\delta(1950)$ &  Mpc &      & mag           & mag     & W m$^{-2}$   &                     & \\
\hline 
\ \ \ [1]     &\ \ \ [2]  &\ \ \ \ \ \ [3] &  [4] & [5]  &\ \ \ [6]   & [7]  & [8]\ \ &\ \ \ \ \ \ \ \ [9]  &\ \ \ \ \ \ [10] \\
\hline 
D~22  &  $\rm 02^h\ 29^m\ 36^s.0 $  & $\rm 38^\circ\ 27'\ 30'' $ & $9.9$  & $2$ & $-14.9$ &$0.09$ & $<-14.49$&{\footnotesize G7}   &{\footnotesize U~2014} \\
D~43  &  $\rm 07^h\ 24^m\ 48^s.0 $  & $\rm 40^\circ\ 52'\ 00'' $ & $4.9$  & $3$ & $-13.9$ &$0.20$ & $<-14.23$&{\footnotesize close to  G6}& {\footnotesize U~3860} \\
D~46  &  $\rm 07^h\ 38^m\ 00^s.0 $  & $\rm 40^\circ\ 14'\ 00'' $ & $4.9$  & $3$ & $-14.7$ &$0.38$ & $<-14.68$&{\footnotesize close to  G6}& {\footnotesize U~3966} \\
D~47  &  $\rm 07^h\ 39^m\ 06^s.0 $  & $\rm 16^\circ\ 55'\ 06'' $ & $2.0$  & $3$ & $-13.4$ &$0.35$ & $ -13.96$&{\footnotesize N~2683}& {\footnotesize U~3974} \\
D~48  &  $\rm 07^h\ 54^m\ 48^s.0 $  & $\rm 58^\circ\ 10'\ 36'' $ & $15.7$ & $3$ & $-16.4$ &$0.59$ & $<-14.41$&{\footnotesize N~2549}&{\footnotesize U~4121} \\
N~2537 & $\rm 08^h\ 09^m\ 42^s.0 $  & $\rm 46^\circ\ 09'\ 00'' $ & $6.4$  & $3$ & $-17.0$ &$0.58$ & $ -12.83$&$^a$ &{\footnotesize U~4274,Mk~86} \\
U~4278 &   			    & 				 &	  &     & $-17.3$ &	  &  &$^{a}$ &{\footnotesize U~4274,Mk~86}  \\	
D~52  &  $\rm 08^h\ 25^m\ 33^s.9 $  & $\rm 41^\circ\ 52'\ 01'' $ & $5.3$  & $3$ & $-13.8$ &$0.53$ & $<-14.24$&{\footnotesize close to G6}&{\footnotesize U~4426} \\
D~63  &  $\rm 09^h\ 36^m\ 00^s.0 $  & $\rm 71^\circ\ 24'\ 54'' $ & $3.4$  & $1$ & $-15.0$ &$-0.09$& $<-14.13$&{\footnotesize M81;L~176}&{\footnotesize U~5139,Ho I} \\
N~2976 & $\rm 09^h\ 47^m\ 15^s.8 $  & $\rm 68^\circ\ 55'\ 00'' $ & $3.4$  & $1$ & $-17.4$ &$0.58$ & $ -12.15$&{\footnotesize M81;L~176}&{\footnotesize U~5221} \\
D~64  &  $\rm 09^h\ 50^m\ 22^s.0 $  & $\rm 31^\circ\ 29'\ 17'' $ & $6.1$  & $2$ & $-14.7$ &$0.15$ & $ -13.90$&{\footnotesize N~2903}$^b$&{\footnotesize U~5272} \\
D~68  &  $\rm 09^h\ 53^m\ 52^s.0 $  & $\rm 29^\circ\ 03'\ 47'' $ & $6.1$  & $2$ & $-14.3$ &$0.23$ & $ -13.90$&{\footnotesize N~2903}&{\footnotesize U~5340} \\
D~73  &  $\rm 10^h\ 09^m\ 34^s.0 $  & $\rm 30^\circ\ 09'\ 03'' $ & $18$   & $2$ & $-16.7$ &$0.62$ & $<-14.25$&{\footnotesize G42}&{\footnotesize U~5478} \\
D~83  &  $\rm 10^h\ 33^m\ 54^s.0 $  & $\rm 31^\circ\ 48'\ 24'' $ & $9 $   & $2$ & $-15.0$ &$0.01$ & $<-14.36$&{\footnotesize G12}&{\footnotesize U~5764}\\
D~87  &  $\rm 10^h\ 46^m\ 18^s.0 $  & $\rm 65^\circ\ 47'\ 36'' $ & $3.4$  & $3$ & $-12.8$ &$0.26$ & $ -14.23$&{\footnotesize M81}&{\footnotesize U~5918,7Zw347} \\
Mk~178 & $\rm 11^h\ 30^m\ 46^s.2 $  & $\rm 49^\circ\ 30'\ 54'' $ & $5.2$  & $3$ & $-15.0$ &  	  & $<-14.20$& &{\footnotesize U~6541} \\
N~3738 & $\rm 11^h\ 33^m\ 00^s.0 $  & $\rm 54^\circ\ 47'\ 00'' $ & $5.2$  & $2$ & $-16.6$ &$0.38$:& $ -13.14$& &{\footnotesize U~6565,Arp 234}\\
D~101 &  $\rm 11^h\ 55^m\ 40^s.8 $  & $\rm 31^\circ\ 30'\ 54'' $ & $7.2$  & $2$ & $-14.7$ &$0.73$ & $ -14.29$&{\footnotesize CVn II(G10)}&{\footnotesize U~6900}\\
D~123 &  $\rm 12^h\ 26^m\ 07^s.9 $  & $\rm 58^\circ\ 19'\ 11'' $ & $11.4$ & $3$ & $-17.5$ &$-0.18$& $ -13.97$&{\footnotesize close to G10}  &{\footnotesize U~7534} \\
Mk~209 & $\rm 12^h\ 23^m\ 51^s.6 $  & $\rm 48^\circ\ 46'\ 12'' $ & $4.9$  & $3$ & $-14.3$ &$0.11$ & $ -13.74$&{\footnotesize L~290?} &{\footnotesize 1Zw36} \\
D~125 &  $\rm 12^h\ 25^m\ 18^s.0 $  & $\rm 43^\circ\ 46'\ 18'' $ & $4.5$  & $2$ & $-15.6$ &$0.40$ & $<-13.98$&{\footnotesize L~290}$^c$ &{\footnotesize U~7577} \\
D~133 &  $\rm 12^h\ 30^m\ 25^s.2 $  & $\rm 31^\circ\ 48'\ 54'' $ & $5.2$  & $2$ & $-15.6$ &$0.44$ & $ -14.11$&{\footnotesize CVn I (G3);L~291}&{\footnotesize U~7698}  \\
D~165 &  $\rm 13^h\ 04^m\ 30^s.0 $  & $\rm 67^\circ\ 58'\ 00'' $ & $4.6$  & $2$ & $-15.8$ &$0.23$ & $<-13.57$&{\footnotesize N~4236}&{\footnotesize U~8201} \\
D~166 &  $\rm 13^h\ 11^m\ 00^s.0 $  & $\rm 36^\circ\ 28'\ 36'' $ & $16$   & $2$ & $-17.6$ &$0.32$ & $ -13.50$&{\footnotesize N~5033}$^d$;{\footnotesize L~334}&{\footnotesize U~8303,Ho VIII} \\
D~168 &  $\rm 13^h\ 12^m\ 12^s.0 $  & $\rm 46^\circ\ 11'\ 00'' $ & $3.5$  & $2$ & $-15.2$ &$0.36$ & $ -13.74$&{\footnotesize Uma/M101;L~347}&{\footnotesize U~8320} \\
D~185 &  $\rm 13^h\ 52^m\ 55^s.2 $  & $\rm 54^\circ\ 09'\ 00'' $ & $6.9$  & $1$ & $-15.6$ &$0.43$ & $ -14.04$&{\footnotesize M101 (G5); L~371}&{\footnotesize U~8837,Ho IV}\\
D~190 &  $\rm 14^h\ 22^m\ 48^s.0 $  & $\rm 44^\circ\ 44'\ 00'' $ & $6 $   & $2$ & $-15.8$ &$0.24$ & $ -14.03$&{\footnotesize (Uma/M101)}&{\footnotesize U~9240} \\
D~216 &  $\rm 23^h\ 26^m\ 06^s.0 $  & $\rm 14^\circ\ 28'\ 00'' $ & $1.0$  & $1$ & $-13.1$ &$0.62$ & $<-14.07$&{\footnotesize Local Group}&{\footnotesize U~12613}$^e$ \\
D~217 &  $\rm 23^h\ 27^m\ 33^s.0 $  & $\rm 40^\circ\ 43'\ 07'' $ & $9.3$  & $3$ & $-17.6$ &$0.58$ &$ -13.66$&{\footnotesize N~7640}&{\footnotesize U~12632} \\
\hline 
\end{tabular} 
\label{phys-sample-tab}
\end{center}
Column Designations: [1] Object name: D = DDO, Mk = Markarian, N = NGC, 
U = UGC; [2, 3] Adopted position for aperture synthesis fringe-stopping 
center; [4] Distances 
compiled from the literature. Stellar distances for {\small DDO\,}63, 
{\small NGC\,}2976, {\small NGC\,}3738, {\small DDO\,}165, 
{\small DDO\,}168 and {\small DDO\,}216 (Karachentsev $\&$ Tikhonov 1994;
Rozanski $\&$ Rowan-Robinson 1994). Group distances from the larger spiral 
group members and based on stellar indicators wherever possible 
(Rowan-Robinson 1988, Rozanski $\&$ Rowan-Robinson 1994) and the blue
Tully-Fisher relation (Kraan-Korteweg et~al. 1988; data from 
de Vaucouleurs et~al. 1991). For a few galaxies recession velocities 
were used, corrected for Local Group peculiar motion and H$_{0}$ = 75 
$\kms$\, Mpc$^{-1}$. [5] Distance quality flag: 1 = reliable distance from 
two or more independent indicators (excluding Hubble expansion); 2 = 
distance from only one reliable indicator; 3 = tentative distance based 
on group membership or recession velocity. [6, 7, 8] Absolute B band 
magnitudes, corrected for Galactic foreground absorption, ($B-V$) colours 
and log(FIR) luminosities all taken from Melisse $\&$ Israel (1994); [9] 
Group membership according to de Vaucouleurs et~al. (1983). Objects 
appearing as a group member in Garcia (1993) are listed as L = LGG. 
[10] Any  other names, mostly UGC.
\\
\\
Notes : {\it a:} {\small NGC~}2537 and {\small UGC~}4278 are a pair of
galaxies; The galaxy {\small NGC~}2537A ($\alpha_{1950}=\rm 8^h10^m09^s.0$,
$\delta_{1950}=\rm 46^\circ8'46''$) could be associated with this pair, but
it is not visible in the HI data. {\it b:} {\small DDO~}64 has a small 
companion (see text); {\it c:} {\small DDO~}125 is a companion of the irregular
galaxy {\small NGC~}4449; {\it d:} {\small DDO~}166 is probably a
companion of {\small NGC~}5033; {\it e:} {\small DDO~}216 is also
called Pegasus dwarf galaxy.
\end{table*} 

%
\begin{table*}
\caption{Galaxies with nearby neighbours}
\begin{center}
\begin{tabular}{|  l |  l |  c   c   c   c |} 
\hline 
Name	 &   Neighbours  &   proj. dist.  &  $\Delta v$  &  type        &   $\Delta m_{\rm B}$ \\ 
         &               &  kpc  & $\kms$ &            &         \\
\hline
\ \ \ \ \ \  [1]   &\ \ \ \ \ \   [2]     &     [3]      &     [4]      &   [5]        &     [6]      \\
\hline
DDO~47   & CGCG~087-033  & $10$  & $10$   &   --       & $2.1$   \\ 
\hline
NGC~2537 & NGC~2537A     & $ 9$  & $ -4$  &  SBc       & $3.7$   \\ 
         & UGC~4278      & $31$  & $116$  &  SBd       & $0.75$  \\ 
\hline
DDO~64   & UGC~5272A     &       &        &            &         \\ 
\hline
DDO~125  & DDO~129       & $25$  & $342$  &  Im        & $0.8$   \\ 
         & NGC~4449      & $48$  & $11$   &  IBm       & $-2.9$  \\ 
         & MCG+07-26-012 & $60$  & $244$  &  --        & $2.2$   \\ 
         & UGC~7690      & $91$  & $341$  &  Im:       & $0.3$   \\ 
	 & MCG+07-26-011 & $105$ & $212$  &  --        & $2.2$   \\ 
         & NGC~4460      & $108$ & $332$  &  SB        & $-0.6$  \\ 
\hline
DDO~166  & UGC~8314      & $50$  & $ -9$  &  Im:       & $3.5$   \\ 
         & NGC~5014      & $100$ & $182$  &  Sa?       & $0.0$   \\ 
         & NGC~5033      & $106$ & $-69$  &  SAc Sy1.9 & $-2.7$  \\ 
\hline
DDO~168  & DDO~167       & $27$  & $-31$  &  Im        & ($4.3$) \\ 
         & UGC~8215      & $89$  & $23$   &  Im        & ($4.3$) \\ 
         & DDO~169       & $97$  & $65$   &  IAm       & $1.9$   \\ 
\hline 
\end{tabular}
\end{center} 
Note: Column 4 lists velocity difference (neighbour - sample dwarf) and
column 6 the B-magnitude difference (neighbour - sample dwarf), i.e.
positive if the neighbour is less luminous.
\label{nearest} 
\end{table*}

%
\begin{table*}
\caption{Isolated galaxies}
\begin{center}
\begin{tabular}{|  l |  l |  c   c   c   c |} 
\hline 
Name 	       &   Neighbours  &   proj. dist.  &  $\Delta v$  &  type        &   $\Delta m_{\rm B}$ \\ 
               &           &   kpc        &  $\kms$ &         &         \\
\hline
\ \ \ \ \ \  [1]   &\ \ \ \ \ \   [2]   &     [3]        &     [4]      &   [5]        &     [6]      \\
\hline
DDO~48         &  NGC~4241   &  $390$           &  $143$         &  Sab    &  $ -0.3$ \\ 
	       &  IC~2209    &  $630$           &  $335$         &  GROUP  &        \\ 
               &  NGC~2460  &  $640 $          &   $358$         &  SAa    &   $-2.5$ \\ 
\hline
DDO~52         &  PGC~22900 &  $326$      &  $353$         &  I?     &   0.3 \\ 
               &  UGC~4278   &  $430$           &  $166$         &  SBd    &  $ -1.9 $\\ 
               &  NGC~2537  &  $460$           &  $50$          &  SBm pec &  $-2.7 $  \\ 
\hline
DDO~83         &  DDO~84     &  $500$           &  $43$          &  Im     &   $-1.5 $\\ 
               &  NGC~3413   &  $520$           &  $59$          &  S0     &   $-2.1 $\\ 
               &  NGC~3274   &  $630$           & $ -49$         &  SABd?  &  $ -2.0 $  \\ 
\hline
DDO~123$^a$    &  (NGC~4605) &  ($740$)         &  ($-580$)      &  (SBc pec) &  ($-3.6$) \\ 
               &  UGC~6931   &  $760$           &  $476$         &  SBm    &   $-0.2$ \\ 
               &  UGC~6926   &  $790$           &  $359$         &  Sdm    &   $1.5$ \\ 
               &  UGC~7544   &  $808$           & $ -13 $        &  dwarf  &   ($2.5$) \\ 
               &  UGC~8146   &  $940$           & $ -53 $        &  Scd    &   $-0.3$ \\ 
\hline
DDO~165$^b$    &  UGC~7748   &  $240$           &  $421$         &  Sdm:   &    $2.5$ \\ 
               &  UGC~7490   &  $370$           &  $430$         &  SAm    &    $0.25$ \\ 
               &  NGC~4236   &  $390$           &  $-37 $        &  SBdm   &    $-2.3$ \\ 
\hline 
\end{tabular}
\end{center} 
Note: Columns as in Table~2; {\it a:} Only the nearest five objects in 
the field are listed here. {\small NGC\,}4605 is ouside the velocity 
range of 500 $\kms$.  {\it b:} {\small DDO\,}165 was added to this list 
because the high relative velocity of {\small UGC\,}7748 suggests that this
galaxy could be a background object.
\label{farthest}
\end{table*}

The majority of the galaxies in the sample (21 out of 29 objects) 
are `true' dwarfs in the sense that their luminosity does not exceed 
$M_{\rm B} = -16$. Four objects ({\small UGC\,}4278, {\small DDO\,}123, 
{\small DDO\,}166 and {\small DDO\,}217) are brighter than $M_{\rm B} = -17$.
All but four galaxies are classified Im (Magellanic irregular). Exceptions
are {\small DDO\,}48 (SBm), {\small NGC\,}2976 (Sd), {\small DDO\,}185
(SBm) and {\small DDO\,}217 (Sm). 
This sample is much larger than the number of dwarf galaxies 
previously studied in HI at comparable resolutions, such as the 
VLA-based studies by Lo et al. (1993) and Puche $\&$ Westpfahl (1993),
and the individual case studies listed in the inventory given by 
Salpeter $\&$ Hoffman (1996).

%
\begin{table}
\caption{WSRT observational parameters}
\begin{tabular}{|  l |  c |  r   r   r   r |} 
\hline 
 Name & obsdate & SP	& $v_c$  & BW   & $\Delta v$ \\ 
       & YYMMDD & m	& $\kms$ & MHz  & $\kms$ \\ 
\hline 
\ \ [1] & [2]   & [3]   & [4]  	 & [5]  & [6] \\ 
\hline 
D~22   & 911006 & 72	&   570  & 2.50 & 4.12 \\ 
D~43   & 861217 & 72	&   355  & 1.25 & 4.12 \\ 
D~46   & 861222 & 72	&   364  & 1.25 & 4.12 \\ 
D~47   & 911019 & 72	&   265  & 2.50 & 4.12 \\ 
D~48   & 881021 & 72	&  1183  & 2.50 & 4.12 \\ 
N~2537 & 911109 & 72	&   450  & 2.50 & 4.12 \\ 
D~52   & 861228 & 72,72	&   392  & 1.25 & 4.12 \\ 
D~63   & 911006 & 72	&   140  & 2.50 & 4.12 \\ 
N~2976 & 870702 & 36,72	&    60  & 2.50 & 8.24 \\ 
D~64   & 840722 & 36	&   520  & 1.25 & 4.12 \\ 
D~68   & 840721 & 36	&   504  & 1.25 & 4.12 \\ 
D~73   & 911108 & 72,72	&  1370  & 2.50 & 4.12 \\ 
D~83   & 911010 & 72	&   585  & 2.50 & 4.12 \\ 
D~87   & 911011 & 72	&   335  & 2.50 & 4.12 \\ 
Mk~178 & 910831 & 72	&   240  & 1.25 & 2.06 \\ 
N~3738 & 911012 & 72	&   225  & 2.50 & 4.12 \\ 
D~101  & 911013 & 72	&   550  & 2.50 & 4.12 \\ 
D~123  & 900910 & 36,72	&   723  & 2.50 & 4.12 \\ 
Mk~209 & 910902 & 72	&   240  & 1.25 & 2.06 \\ 
D~125  & 911017 & 72	&   195  & 2.50 & 4.12 \\ 
D~133  & 910906 & 72	&   331  & 1.25 & 2.06 \\ 
D~165  & 870507 & 54	&    33  & 1.25 & 4.12 \\ 
D~166  & 881016 & 72	&  1009  & 2.50 & 4.12 \\ 
D~168  & 911019 & 72	&   200  & 2.50 & 4.12 \\ 
D~185  & 910907 & 72	&   141  & 1.25 & 2.06 \\ 
D~190  & 861228 & 72	&   153  & 1.25 & 4.12 \\ 
D~216  & 911020 & 72	&  -180  & 2.50 & 4.12 \\ 
D~217  & 840716 & 36	&   424  & 1.25 & 4.12 \\ 
\hline 
\end{tabular} 
\label{obs-sample-tab}
\end{table}

Interactions may have a profound effect on structure and evolution of 
galaxies, especially dwarf galaxies. We have therefore searched the 
NED\footnote{The NASA/IPAC Extragalactic Database (NED) is operated by 
the Jet propulsion Laboratory, California Institute of Technology, under 
contract with the National Aeronautics and Space Administration} for 
galaxies within $5^\circ$ of the objects in our sample. This field of 
view corresponds to a linear size of 900 kpc at 5 Mpc, the median 
distance of the sample galaxies.  
Most of the sample dwarf galaxies are in groups, but not necessarily 
close to massive members. As satellites of the Galaxy or of M31 are 
usually found within a radius of 100-200 kpc and with a relative velocity
of a few hundred $\kms$ (Lynden-Bell, 1994), we have first listed
dwarfs with such relatively nearby neighbours in Table~\ref{nearest}. 
The predominance of positive velocity differences reflects a bias 
caused by the large velocity window as compared to the typical systemic 
velocity. The resulting large volume effect introduces more interlopers 
at higher velocities.

Six dwarf galaxies have such nearby companions. {\small NGC\,}2537, 
and {\small UGC\,}4278 are close enough in space and velocity to 
include in a single WSRT field of view. The other companion,
{\small NGC\,}2537A is not visible in the WSRT field. Neither is 
is {\small CGCG\,}087-033, a companion of {\small DDO\,}47,
discovered by Walter $\&$ Brinks (2001). The dwarf companion 
of {\small DDO\,}64 (={\small UGC\,}5272) was first described by Hopp 
$\&$ Schulte-Ladbeck (1991) and is referred to as {\small UGC\,}5272B 
following these authors.  It can likewise be included with {\small
DDO\,}64 in a single WSRT field. Although there are no large nearby
galaxies, the dwarfs {\small UGC\,}5209, {\small UGC\,}5186, 
{\small DDO\,}68, {\small UGC\,}5427 and {\small UGC\,}5464 occur
not far away (projected distance less than 500 kpc and velocites within 
40 $\kms$). {\small DDO\,}125 is located near the edge of the large HI 
halo of the irregular galaxy {\small NGC\,}4449 (Bajaja et al. 1994).
{\small DDO\,}166 is a member of the {\small NGC\,}5033 group. Both
{\small NGC\,}5033 and {\small UGC\,}8314 can be included in the same
WSRT observation.  Finally, {\small DDO\,}168 is not far from the 
low-luminosity dwarf {\small DDO\,}167. {\small NGC\,}5023 is at a 
projected distance of 117 kpc and a relative velocity 213 $\kms$.

The most {\it isolated dwarfs} in the sample are {\small DDO\,}48d, 
{\small DDO\,}52, {\small DDO\,}83, {\small DDO\,}123, {\small DDO\,}165 
and {\small DDO\,}216 (Table~\ref{farthest}). The projected distance 
between {\small DDO\,}52 and its nearest known neighbour, {\small PGC~}22900, 
is 326 kpc, and its large velocity difference suggests that {\small PGC~}22900 
is an unrelated background galaxy. The {\small NGC\,}2537/{\small UGC\,}4278 
pair is part of the present sample; this pair is relatively far from
{\small DDO\,}52, but at nearly the same velocity. The nearest neighbour 
of {\small DDO\,}83 is {\small DDO\,}84, at a very similar velocity but
at a projected distance of 500 kpc, just as {\small NGC\,}3413 and 
{\small NGC\,}3274. {\small DDO\,}123 appears to be the most isolated galaxy 
in the sample. The data in Table~\ref{farthest} suggest that 
{\small DDO\,}123 is associated with {\small UGC\,}7544 and 
{\small UGC\,}8146, so that the nearest neighbour is a dwarf galaxy 
at least 800 kpc distant. Four of the `isolated' objects in 
Table~\ref{farthest} are accompanied by late-type galaxies with radial 
velocities not differing by more than 50 $\kms$. As the number density of 
galaxies along the velocity axis is low, it is highly unlikely that these
galaxies are unrelated.

\section{Observations and reduction}              

All observations discussed in this paper were made with the Westerbork 
Synthesis Radio Telescope (WSRT) between 1984 en 1991. At the
observing fequency of 1.4 GHz, the primary beam is 40$'$ (FWHM),
and the synthesized beam is $13''\times\,13''$sin$\delta$. For a
galaxy at distance $d_{\rm Mpc}$, this beamsize corresponds to a linear
resolution of $63 \cdot d_{\rm Mpc}$ pc. For the nearest dwarf galaxies 
in our sample this is approximately the size of the Orion starforming 
complex. The largest scales which can be studied are mostly determined 
by the sensitivity to low-column-density HI in the outer regions; the
sensitivity of a 12 hour observation with the WSRT is about $10^{20}$ 
HI atoms $\rm cm^{-2}$ or 0.8 $\rm M_\odot\ pc^{-2}$. Each 
galaxy was observed for a full 12-hour period, resulting in a map 
with the first grating response at a radius of 10$'$. Four galaxies
were observed for 2 x 12 hours; for {\small NGC\,}2976 and {\small 
DDO\,}123 the first grating response is at a radius of 20$'$. In 
general, the shortest baseline sampled was 72 m (341 wavelengths). 
Five galaxies were observed with a shortest spacing of 36 m (171 
wavelengths) and one with a shortest spacing of 54 m. The lack of short 
spacings causes a depression of the map zerolevel proportional to source 
strength and a dilution of the interferometer response to extended 
structures. However, the limited overall extent of the sample galaxies, 
and the limited extent of emission in any particular velocity channel 
served to greatly minimize any interferometric dilution. Flux and 
phase were calibrated by using the strong radio sources 3C48, 3C147 
and 3C286. Additional information in Table \ref{obs-sample-tab}, where 
we list for each galaxy observation the date, the shortest spacing SP 
in m, the central velocity $v_{c}$ in $\kms$, the total bandwidth BW in 
MHz, and the channel separation $\Delta v$ in $\kms$. All data were 
Hanning-tapered, so that the velocity resolution is similar to a 
gaussian with a {\small FWHM} equal to $2 \cdot \Delta v$.

Standard gain and phase calibrations were applied by the WSRT 
reduction group in Dwingeloo (cf. Bos et~al. 1981). The {\small 
UV} data were subsequently imported into the NEWSTAR reduction 
package for inspection and Fourier transformation. Data points of
poor quality were excluded from the analysis in all frequency
channels, in order to keep identical {\small UV}-coverage for all 
frequency channels. In each channel, maps were constructed with 
resolutions of $13''.5$, $27''$ and, for the larger objects,
$60''$ ({\small FWHM}) resolution. At decreasing resolutions, the 
circular gaussian weight functions used progressively circularize 
the synthesized beams, because the longer projected baselines 
are suppressed more strongly than the shorter projected baselines. 
For convenience in defining `clean' areas at later reduction stages,
we produced all output maps with identical pixel sizes of $5'' \times \,
5''$, fully sampling the high-resolution beam and oversampling 
the low-resolution beam by a factor of two. 
Separately produced continuum maps (see below) were sampled 
with $5'', 10''$ and $20''$ pixels respectively. For each datacube, 
we produced five antenna patterns distributed evenly over the whole 
frequency band. This was sufficient for the expected 0.18$\%$ size 
variation of the synthesized beam over a 2.5 MHz passband at 1420 MHz. 
  
%
\begin{figure*}
\mbox{
\begin{minipage}[b]{7cm}
\resizebox{7cm}{!}{\includegraphics[angle=-90]{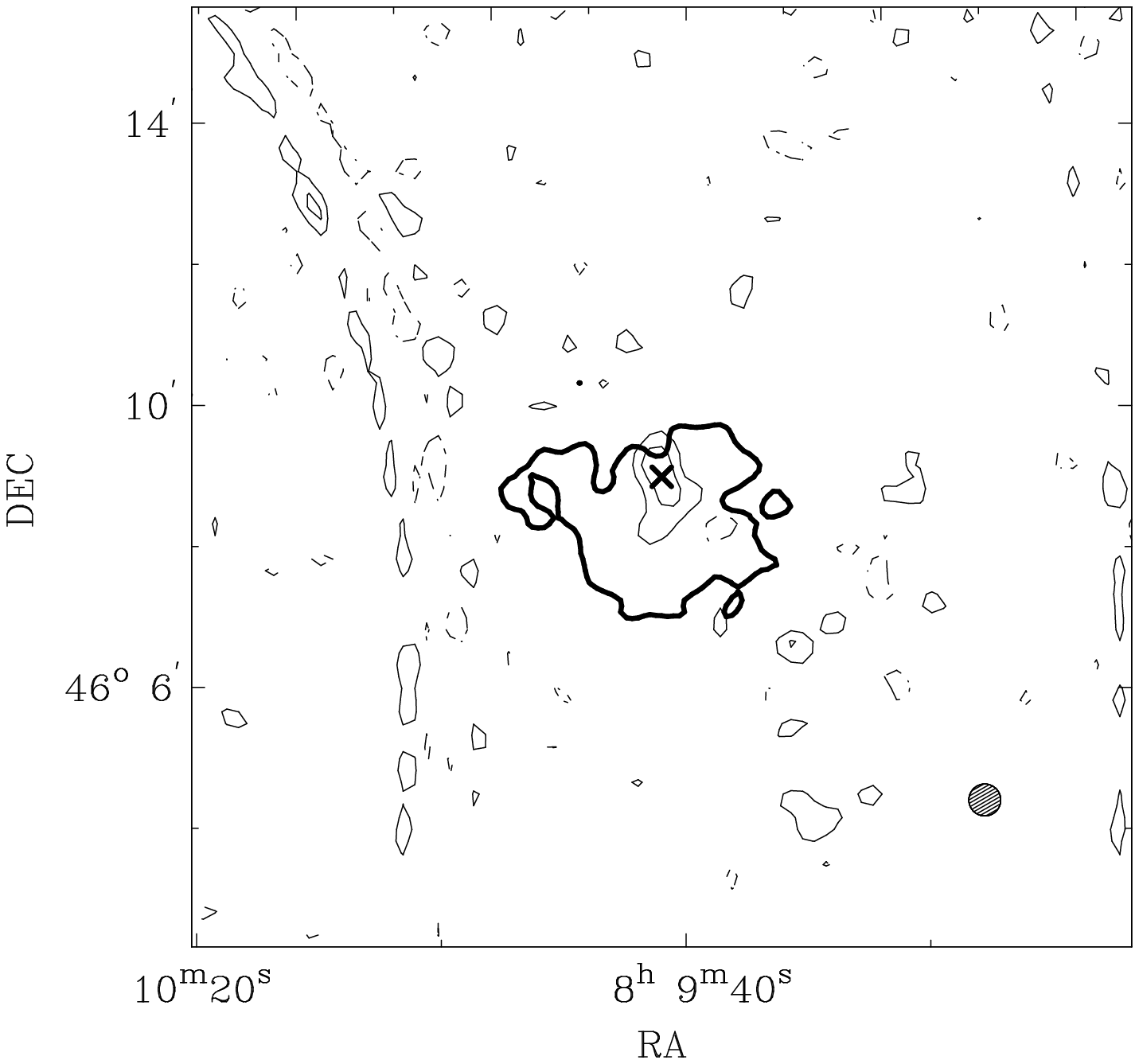}}
\end{minipage}
\begin{minipage}[b]{7cm}
\resizebox{7cm}{!}{\includegraphics[angle=-90]{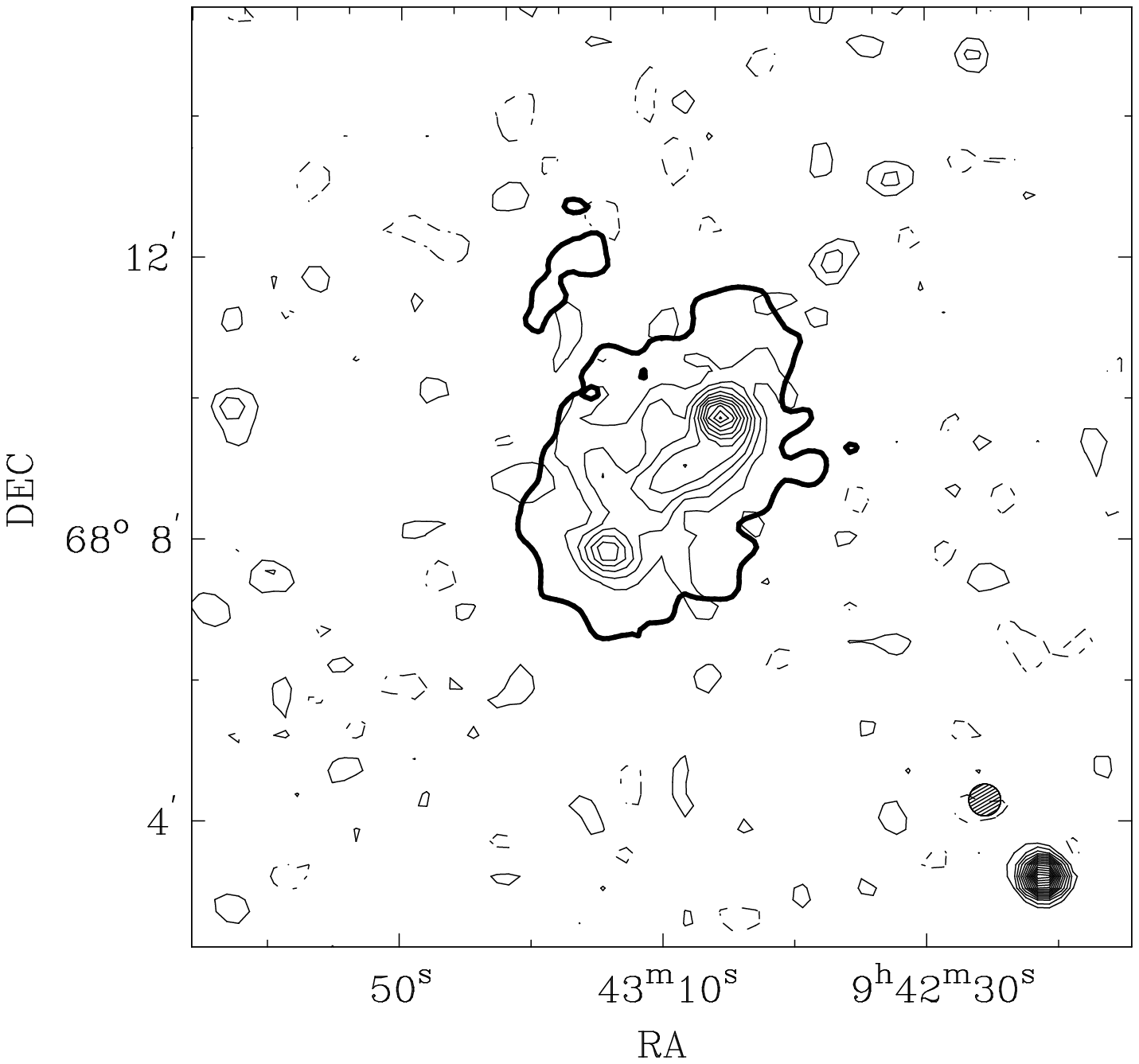}}
\end{minipage}
}
\mbox{
\begin{minipage}[b]{7cm}
\resizebox{7cm}{!}{\includegraphics[angle=-90]{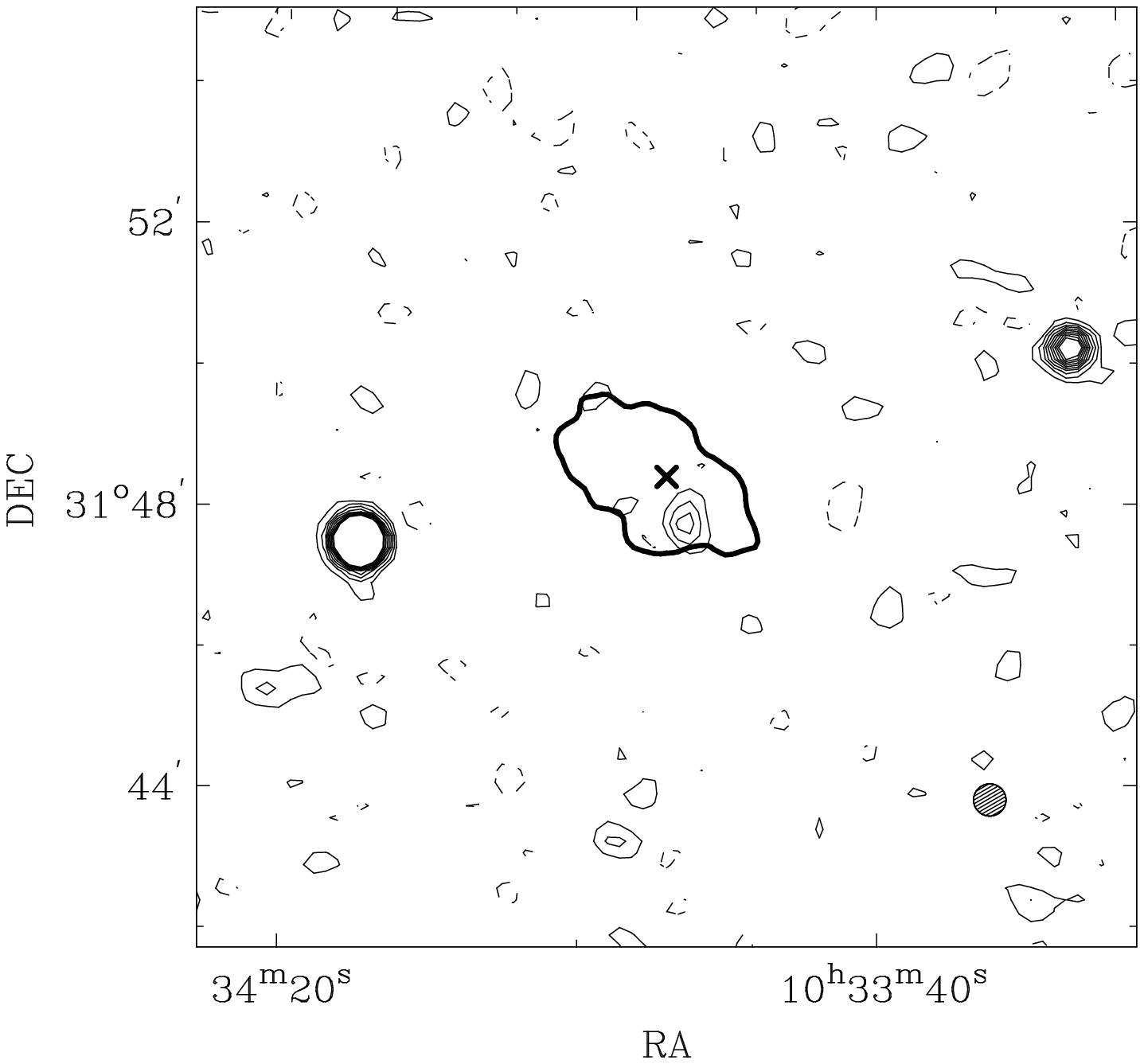}}
\end{minipage}
\begin{minipage}[b]{7cm}
\resizebox{7cm}{!}{\includegraphics[angle=-90]{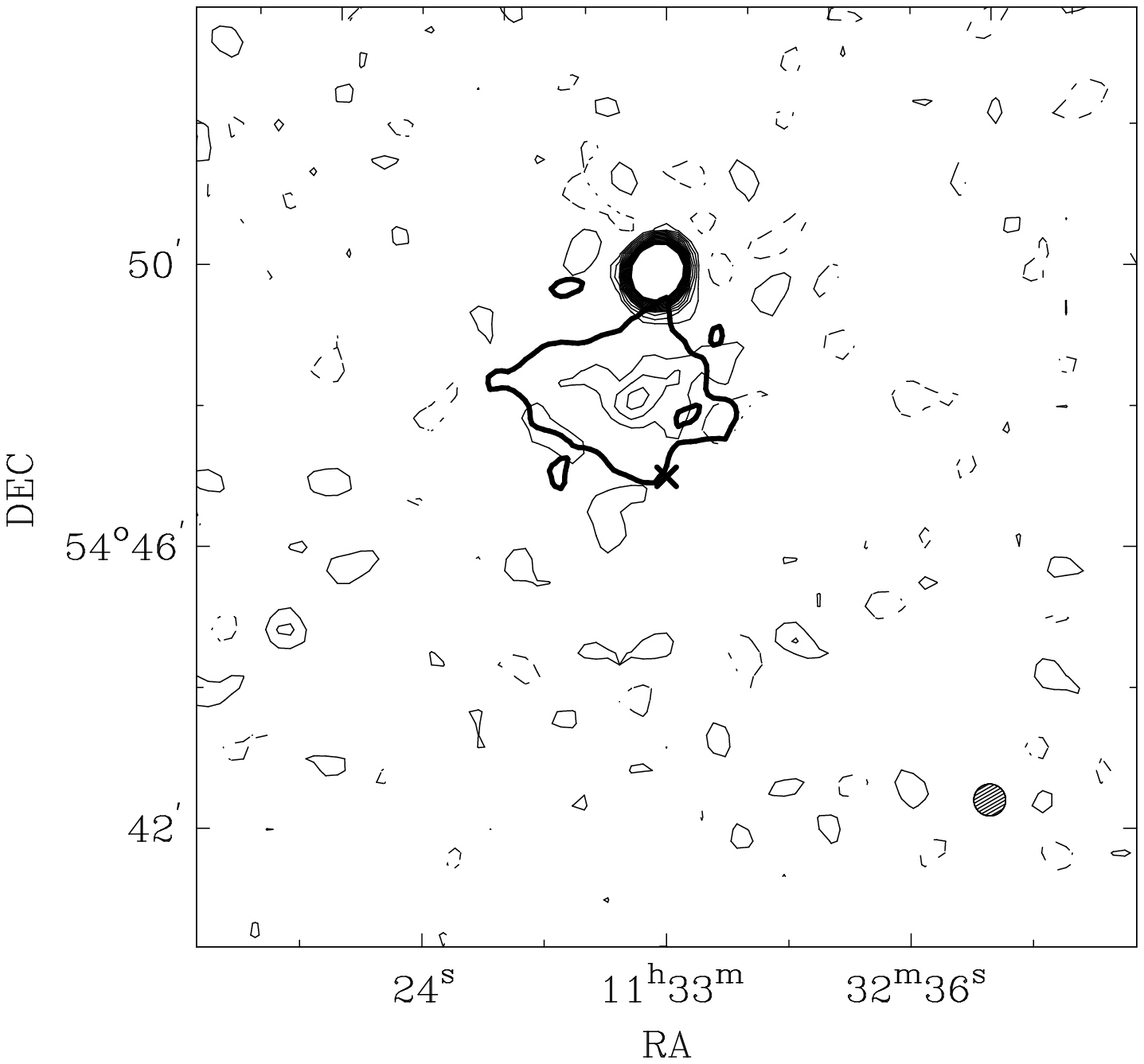}}
\end{minipage}
}
\caption{\small Clockwise from top left: continuum maps of NGC~2537, 
NGC~2976, DDO~83 and NGC~3738 at 27$''$ resolution. The thin contours 
give the 1.4 GHz continuum at the levels $-2$, $2$, $4$, $6$ times 
the rms noise of $0.7$, $0.3$, $0.5$ and $0.6$ mJy per beam area for
NGC~2537, NGC~2976, DDO~83 and NGC~3738 respectively.  The thick contour
is the $3\cdot 10^{20}\rm\, cm^{-2}$ HI column density contour.
Crosses indicate the fringe stopping center, which is outside the
field of view for NGC~2976.
\vspace{0.3cm}
}
\label{cont-sample}
\end{figure*}

%
\vspace{0.1cm}
\begin{table}
\caption{Dwarf galaxies 1.4 GHz continuum emission}
\begin{tabular}{|  l  |  r  r ||  l  |  r  r |} 
\hline 
 Object  &$\rm BW_c$ & $\rm S_{1.4}$ &  Object  &$\rm BW_c$ & $\rm S_{1.4}$ \\
    & MHz & mJy &    & MHz & mJy \\
\hline 
\ \ [1]   & [2] \ \ & [3] \ &\ \ [1] & [2] \ \  & [3] \ \\
\hline 
DDO~22   & 1.58 & $<$ 3.6  & Mk~178   & 0.81 & $<$ 21:  \\ 
DDO~43   & 0.51 & $<$ 4.9  & NGC~3738 & 1.31 & 13$\pm$2 \\
DDO~46   & 0.41 & $<$ 4.3  & DDO~101  & 1.66 & - \\
DDO~47   & 1.29 & $<$ 4.3  & DDO~123  & 1.31 & $<$ 4.0 \\
DDO~48   & 1.23 & $<$ 6.8  & Mk~209   & 0.59 & $<$ 5.4 \\
NGC~2537 & 1.54 & 12$\pm$2 & DDO~125  & 1.60 & $<$ 10  \\
DDO~52   & 0.37 & $<$ 7.3  & DDO~133  & 0.55 & $<$ 7.1 \\
DDO~63   & 1.54 & $<$ 5.9  & DDO~165  & 0.61 & $<$ 5.7 \\
NGC~2976 & 1.17 & 65$\pm$5 & DDO~166  & 1.60 & $<$ 7.0 \\
DDO~64   & 0.41 & $<$ 8.8  & DDO~168  & 1.27 & $<$ 5.5 \\
DDO~68   & 0.39 & $<$ 15 \ \ & DDO~185  & 0.34 & $<$ 11 \ \ \\
DDO~73   & 1.58 & $<$ 5.2  & DDO~190  & 0.33 & $<$ 7.1 \\
DDO~83   & 1.33 & 4$\pm$1  & DDO~216  & 1.35 & $<$ 6.2 \\
DDO~87   & 1.56 & $<$ 7.2  & DDO~217  & 0.39 & $<$ 9.0 \\
\hline 
\end{tabular} 
\label{cont-sample-tab}
\end{table}

After Fourier transformation, the raw datacubes were imported into the
{\small GIPSY} package and searched for line emission from the object 
and the Galaxy foreground. Excluding noisy channel maps at the
edges of the frequency band and channels containing line emission,
we fitted a first order polynomial to the remaining continuum channels.
For each galaxy we took the lowest-resolution datacube, and marked 
in each continuum-subtracted channel map the area containing line 
emission. These areas were stored and used for all datacubes as input 
search areas for the clean algorithm. The clean areas thus manually 
defined were  larger than those defined through clipping, and have 
smooth boundaries. Within the search areas, the clean algorithm 
searched for positive as well as negative components to a level of half 
the r.m.s. noise value as determined from empty channels. The 
components were restored onto the map with gaussian beams fitted
to the antenna pattern at the center of the frequency band.

%
\begin{figure*}
\centerline{\resizebox{15cm}{!}{\includegraphics[angle=0]{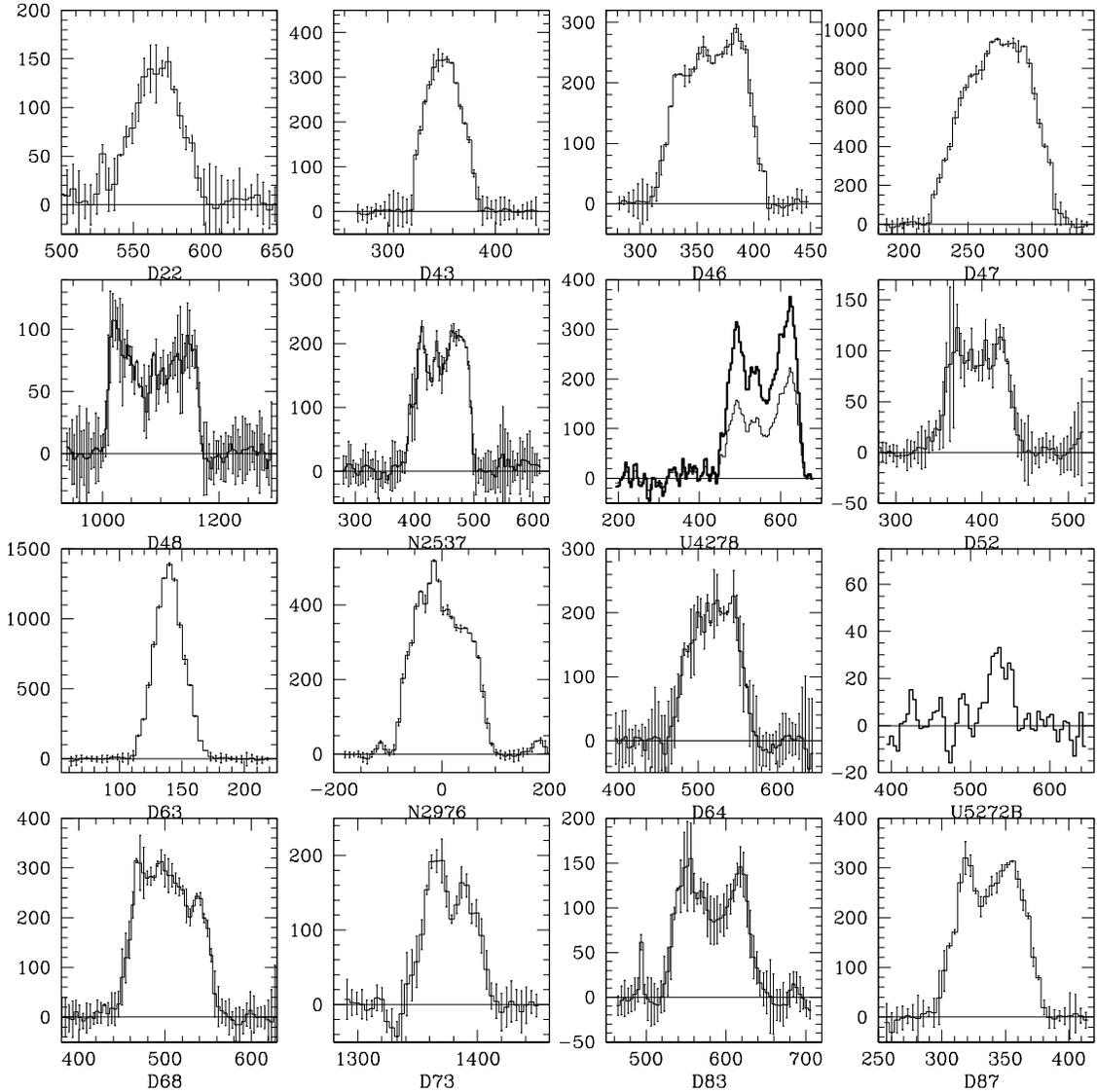}}}
\caption{Integrated (global) line profiles of the dwarf galaxy sample.
The ordinate is heliocentric velocity in $\kms$, the abscissa is 
flux-density in mJy, corrected for primary beam attenuation. For 
{\small UGC\,}4278 the uncorrected flux is also shown. The narrow 
peak in the line profile of {\small NGC\,}2976 near $v_{\rm hel}=0$ is 
caused by emission from the Galactic foreground.}
\label{prof-sample}
\end{figure*}

\begin{figure*}
\addtocounter{figure}{-1}
\centerline{\resizebox{15cm}{!}{\includegraphics[angle=0]{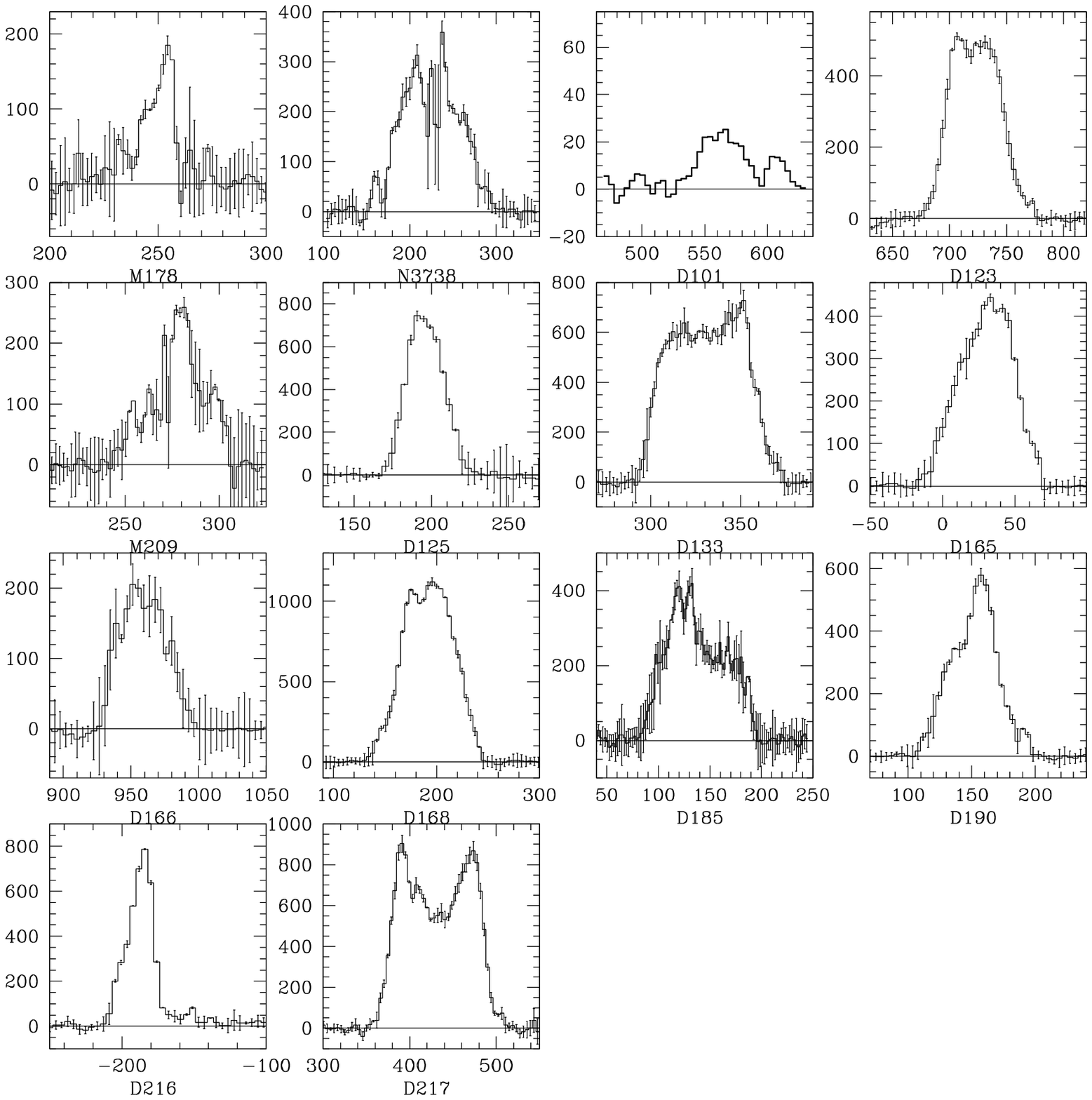}}}
\caption{continued}
\end{figure*}

We added the line-free channels at either side of the passband to
make a `broad band' continuum map. These maps were made in separate 
runs because they must be bigger in order to allow us to `clean'
bright continuum (background) sources far from the map center. 

\section{Results}

\subsection{Radio continuum emission from dwarf galaxies}

Although only a limited bandwidth was available in the continuum,
the large collecting area of the WSRT and the long integration times 
allowed determination of continuum flux-densities of the sample galaxies 
at sensitivities of a few mJy, comparable to those of published 
single-dish surveys (Klein et~al. 1986, Klein 1986, Altschuler et~al. 
1984), but unlike these unhampered by background confusion (cf.
Hoeppe et al. 1994). The sensitivity of our continuum maps is 
determined by the effective bandwidth used, i.e on the line-free
part of the original bandwidth. The effective continuum bandwidths 
are given in Table~\ref{cont-sample-tab}. 

We integrated fluxes inside $2'$ by $2'$ boxes centered on each
galaxy's position, and on at least five map positions judged to be
free of continuum emission in order to determine the local noise 
level. Results are listed in Table~\ref{cont-sample-tab}. Upper limits 
are three times the r.m.s. noise. The high upper limits for Mk~178 and 
{\small DDO\,}68 are due to residual sidelobes of unrelated strong 
continuum sources in the field. For the same reason no value is listed 
for {\small DDO\,}101. Only four out of 28 galaxies (15$\%$) were 
unequivocally detected in the continuum: {\small NGC\,}2537, {\small 
NGC\,}2976, {\small NGC\,}3738 and {\small DDO\,}83 of which maps 
are shown in Fig.~\ref{cont-sample}. This detection rate is similar
to those yielded by single dish continuum surveys. It may suffer from a
positive bias because blue compact dwarf galaxies, which tend to be
more luminous at radio wavelengths (Klein et~al. 1984), are 
overrepresented in our sample.

The best continuum detection (13$\sigma$) is that of 
{\small NGC\,}2976. Our vale S$_{1.4}$ = 64.5 mJy is consistent
with Condon's (1987) value S$_{1.5}$ = 50.8 mJy. The higher
frequency determination S$_{4.85}$ = 29$\pm$5 mJy by Gregory
$\&$ Condon (1991) suggests an overall nonthermal spectral index
$\alpha$ = --0.8 (S$\propto\nu^{\alpha}$). The map shows two 
bright compact sources, almost coincident with the two regions 
of highest HI column density (see below) and connected by more 
diffuse emission. Within a $30''$ aperture, these sources have
flux densities, uncorrected for underlying extended emission,
of $S_{1.4} = 4 \pm 1\ \rm mJy$ (SE: $\alpha=\rm 9^h43^m18^s$,
$\delta=68^\circ7'50''$) and $S_{1.4}= 6 \pm 1.5\ \rm mJy$ (NW:
$\alpha=\rm 9^h43^m2^s$, $\delta= 68^\circ9'50''$), where the 
errors reflect the uncertainty in the definition of the emitting 
regions. We have verified that these radio continuum sources 
correspond precisely with major H$\alpha$ emission regions in the 
galaxy (cf. Stil 1999). Their emission is thus almost certainly 
thermal. The two objects then have excitation parameters $u \approx$
500 resp. 700 pc cm$^{-2}$, corresponding to the Lyman continuum 
photon output of 270 resp. 400 O6 stars, i.e. to major star 
formation regions.

The other three detections are less strong. A just resolved
source coincides with strong H$\alpha$ emission in the main
body of {\small NGC\,}2537 (Stil 1999). Higher-frequency 
flux-densities $S_{4.8} = 18 \pm 4\ \rm mJy$ at 6.3 cm and 
$S_{25}=11 \pm 4\ \rm mJy$ (Klein et~al. 1984) agree well with 
our value in Table~\ref{cont-sample-tab} and suggest a flat 
radio spectrum, i.e. optically thin thermal emission as suggested 
by the strong H$\alpha$ emission. The greater distance of 
{\small NGC\,}2537, and the observed flux-density suggest a
total excitation parameter $u \approx$ 1050 pc cm$^{-2}$, 
corresponding to the output of some 2800 O6 stars. This
suggests a major burst of star formation within the 
kiloparsec-sized region. It is tempting to speculate that
the companion UGC~4278 bears responsibility for this. 

{\small NGC\,}3738 also has a somewhat extended radio
continuum source coincident with the optical galaxy. No other
radio continuum information is available, because of confusion
with the nearby, relatively strong point source 1132+5450
(whose contribution is, of course, not included in our 
determination).  We have searched for HI in absorption
against this source but found none down to a (1 $\sigma$) level 
of $10\%$ ($\tau < 0.1$). However, this upper limit does not 
provide strong constraints on the HI size of {\small NGC\,}3738.

Finally, a faint and somewhat marginal source appears in the map 
of {\small DDO\,}83. This source was accepted as a detection because 
it is resolved and occurs within the HI emission boundary. Optically,
nothing is evident at its position. An even more marginal detection 
$S_{4.8}=3.0 \pm 2.2$ (Klein et~al. 1992) is at least consistent 
with our result. The Effelsberg beamsize of 2.5$'$ was small enough 
to exclude confusion by other sources in the field. 

\subsection{Serendipitous objects}

%
\begin{figure}
\begin{minipage}[b]{9cm}
\resizebox{8.5cm}{!}{\includegraphics[angle=0]{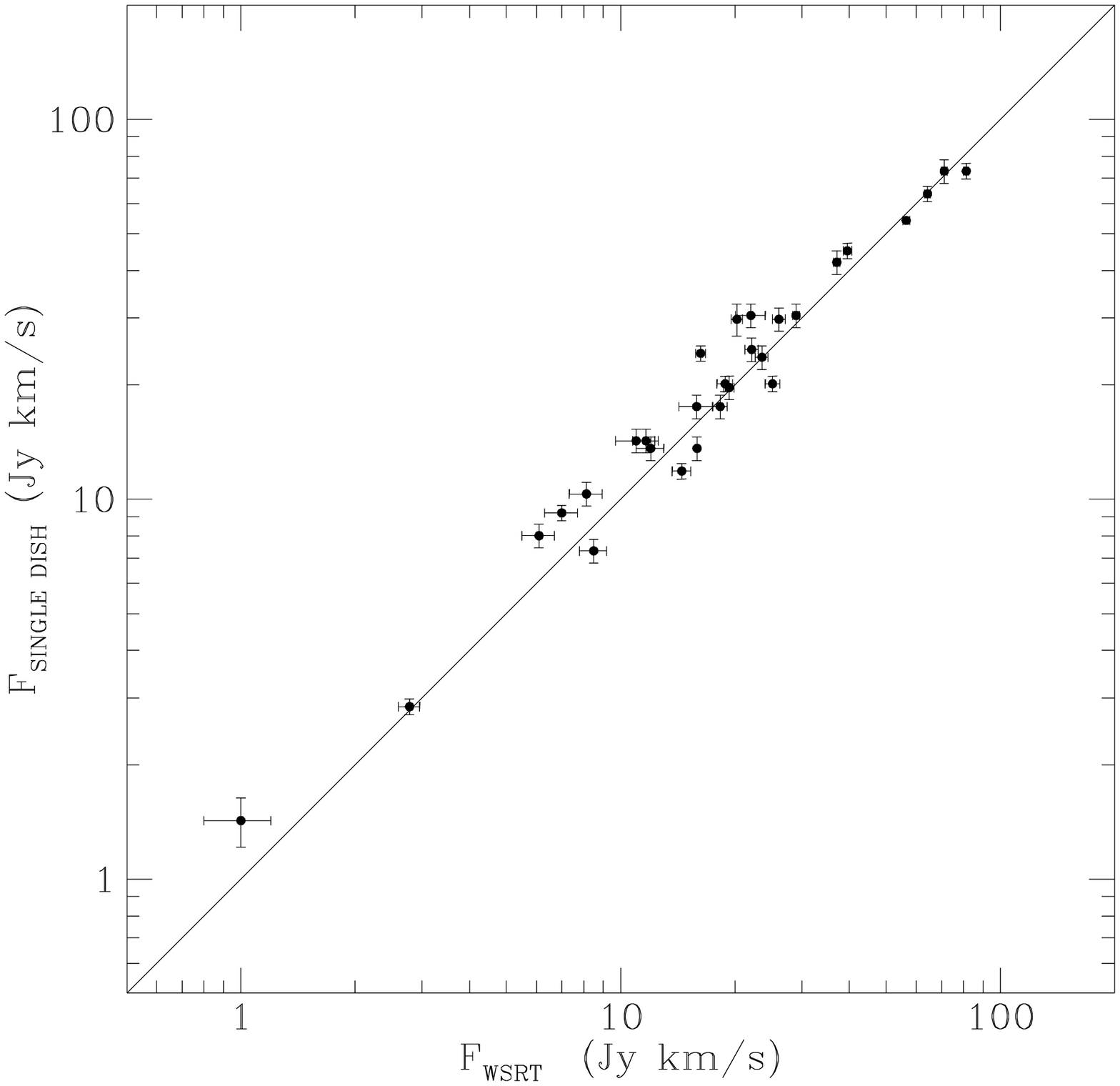}}
\end{minipage}
\caption{HI fluxes from the present WSRT sample
compared with single dish fluxes from Bottinelli et al. (1990). The solid
line connects points with equal flux. } 
\label{fluxflux}
\end{figure}

In a few cases, other galaxies are included in the same WSRT datacube
as the target galaxy. This is the case for {\small UGC\,}4278, companion 
of {\small NGC\,}2537, which occurs at a large distance to the field
center, causing a primary beam attenuation $\approx 50\%$ changing 
significantly over the galaxy. The other companion, {\small NGC\,}2537A 
is not visible in the Westerbork data. Likewise, {\small UGC\,}5272B 
is clearly visible in the {\small DDO\,}64 channel maps at 
$\alpha=\rm 9^h47^m24^s$, $\delta=\rm31^\circ41' 25''$ in the velocity 
range $528<v_{\rm hel}<541\ \kms$.  It can also be seen in the HI 
column density map of {\small DDO\,}64 in Fig.~\ref{NHI-sample}, but 
there it is no stronger than the noise feature to the north of 
{\small DDO\,}64. An HI line profile of {\small UGC\,}5272B is included
in Fig.~\ref{prof-sample}. HI emission from {\small DDO\,}166's
companion {\small NGC\,}5033 is discernable only at the noiselevel;
although {\small UGC\,}8314 appears in the same field of view, it 
cannot be studied with the present data. Although additional uncatalogued 
dwarf galaxies may occur in the vicinity of the sample galaxies, no 
previously unknown objects have been found in the present survey.

\subsection{Integrated HI emission}     

%
\begin{table}
\caption{\label{line-tab} Global HI parameters of dwarf galaxies }
\begin{center}
\noindent
\begin{tabular}{|  l  |  r   r |  r  r  r | } 
\hline 
Name  & $v_{\rm sys}$\ \  & $\int$Sdv \ \ \ \ \  & $M_{\rm HI}$ &  
{$M_{\rm HI}\over{L_{\rm B}}$} & {$M_{\rm HI}\over{L_{\rm FIR}}$}   \\ 
      & $\kms$  & $\rm Jy\ \kms$ & $10^8 \Msun$ 
& \multicolumn{2}{c}{{$\Msun\over{\Lsun}$}} \\
\hline 
\ [1] &  [2] \ \ \  & [3]\ \ \ \ \ \ \ &[4]\ \ & [5] & [6] \\
\hline
D~22  &    564  &   6.1    $\pm$ 0.6  &  1.41 & 1.0 &$>$14 \\ 
D~43  &    353  &  14.5    $\pm$ 0.8  &  0.82 & 1.4 &$>$19 \\
D~46  &    364  &  19.3    $\pm$ 0.6  &  1.09 & 0.9 &$>$72 \\ 
D~47  &    274  &  64.3    $\pm$ 1.2  &  0.61 & 1.7 &   47 \\ 
D~48  &   1084  &  11.7    $\pm$ 0.9  &  6.77 & 1.2 &$>$23 \\
N~2537&    443  &  18.8    $\pm$ 0.9  &  1.82 & 0.2 &  1.0 \\ 
U~4278&    558  &  44.9    $\pm$ 2.0  &  4.33 & 0.3 & ...  \\ 
D~52  &    395  &   8.1    $\pm$ 0.8  &  0.54 & 1.0 &$>$10 \\ 
D~63  &    139  &  39.5    $\pm$ 1.0  &  1.08 & 0.7 &$>$40 \\
N~2976&      3  &  56.5    $\pm$ 1.2  &  1.54 & 0.1 &  0.6 \\
D~64$^a$&  515  &  14.9    $\pm$ 1.6  &  1.30 & 1.1 &  9.2 \\
U~5272B&   537  &   1.0    $\pm$ 0.1  &  0.09 & ... & ...  \\ 
D~68  &    500  &  26.1    $\pm$ 1.0  &  2.29 & 2.8 &   17 \\ 
D~73  &   1376  &   8.5    $\pm$ 0.7  &  6.48 & 0.9 &$>$11 \\ 
D~83  &    582  &  12.0    $\pm$ 1.0  &  2.29 & 1.4 &$>$21 \\
D~87  &    340  &  18.3    $\pm$ 0.8  &  0.50 & 2.4 &   25 \\ 
Mk 178 &  250   &   2.8    $\pm$ 0.5  &  0.18 & 0.1 & $>$3 \\ 
N~3738&    225  &  22.0    $\pm$ 2.0  &  1.40 & 0.2 &  2.4 \\
D~101 &    498  &   1.0    $\pm$ 0.2  &  0.12 & 0.1 &  1.5 \\
D~123 &    723  &  29.0    $\pm$ 0.6  &  8.91 & 0.6 &   21 \\ 
Mk 209 &   280  &   7.0    $\pm$ 0.7  &  0.40 & 0.5 &  3.1 \\ 
D~125 &    196  &  22.2    $\pm$ 0.9  &  1.06 & 0.4 &$>$16 \\ 
D~133 &    331  &  37.0    $\pm$ 0.8  &  2.36 & 0.9 &   38 \\
D~165 &     29  &  20.2    $\pm$ 0.7  &  1.01 & 0.3 & $>$5 \\ 
D~166 &    961  &  11.0    $\pm$ 1.3  &  6.64 & 0.4 &  2.8 \\ 
D~168 &    191  &  71.1    $\pm$ 1.3  &  2.05 & 1.1 &   31 \\ 
D~185 &    136  &  25.1    $\pm$ 1.1  &  2.82 & 1.0 &   22 \\
D~190 &    151  &  23.6    $\pm$ 0.9  &  2.00 & 0.6 &   20 \\ 
D~216 & $-189$  &  16.3    $\pm$ 0.5  &  0.04 & 0.2 &$<$13 \\
D~217 &    433  &  81.3    $\pm$ 1.3  & 16.56 & 1.0 &   29 \\
\hline 
\end{tabular} 
\label{line-tab}
\end{center}
Note : {\it a :} Flux does not include UGC~5272B.
\end{table}

For each object, we extracted the integrated HI line profile from its 
low resolution datacube which provided the best signal-to-noise ratio. 
In each channel map, we determined the cumulative flux as a function 
of concentric box area over the size range of 1 to 20 arcmin. This
procedure yielded for each velocity channel an accurate flux with 
minimum noise contribution. It also provided a good check on possible 
systematic errors such as residual zero offsets left by imperfect 
cleaning. We found such effects to be small compared to the uncertainty 
in the flux calibration, except for {\small NGC\,}3738 where we had 
to apply a relatively large flux correction factor of 1.4. 

As we have already mentioned, the lack of the shortest spacings may cause 
us to underestimate the total flux contained in the interferometer map.
In order to verify the importance of this effect, we have compared
the WSRT flux-integrals just determined with flux-integrals based on
single dish measurements.  In Fig.~\ref{fluxflux} we have plotted the
values listed in the large compilation by Bottinelli et al. (1990)
as a function of the WSRT values. The errorbars do not take into account 
the (systematic) flux calibration uncertainty, which is about of 10$\%$. 
There is good agreement is good, but the WSRT fluxes tend to be lower
by factors up to 1.25. However, as an anonymous referee has pointed out, 
the values listed by Bottinelli et al. (1990) represent a very inhomogeneous 
sample, incorporating corrections for assumed HI extent and corrections
for telescope-deptendent HI flux scales. The magnitude of and uncertainty
in these corrections may easily be comparable to the difference in
integral values noted above. 

%
\begin{figure*}[t]
\unitlength1cm
\begin{minipage}[t]{5.7 cm}
\resizebox{5.7cm}{!}{\includegraphics[angle=-90,clip=true]{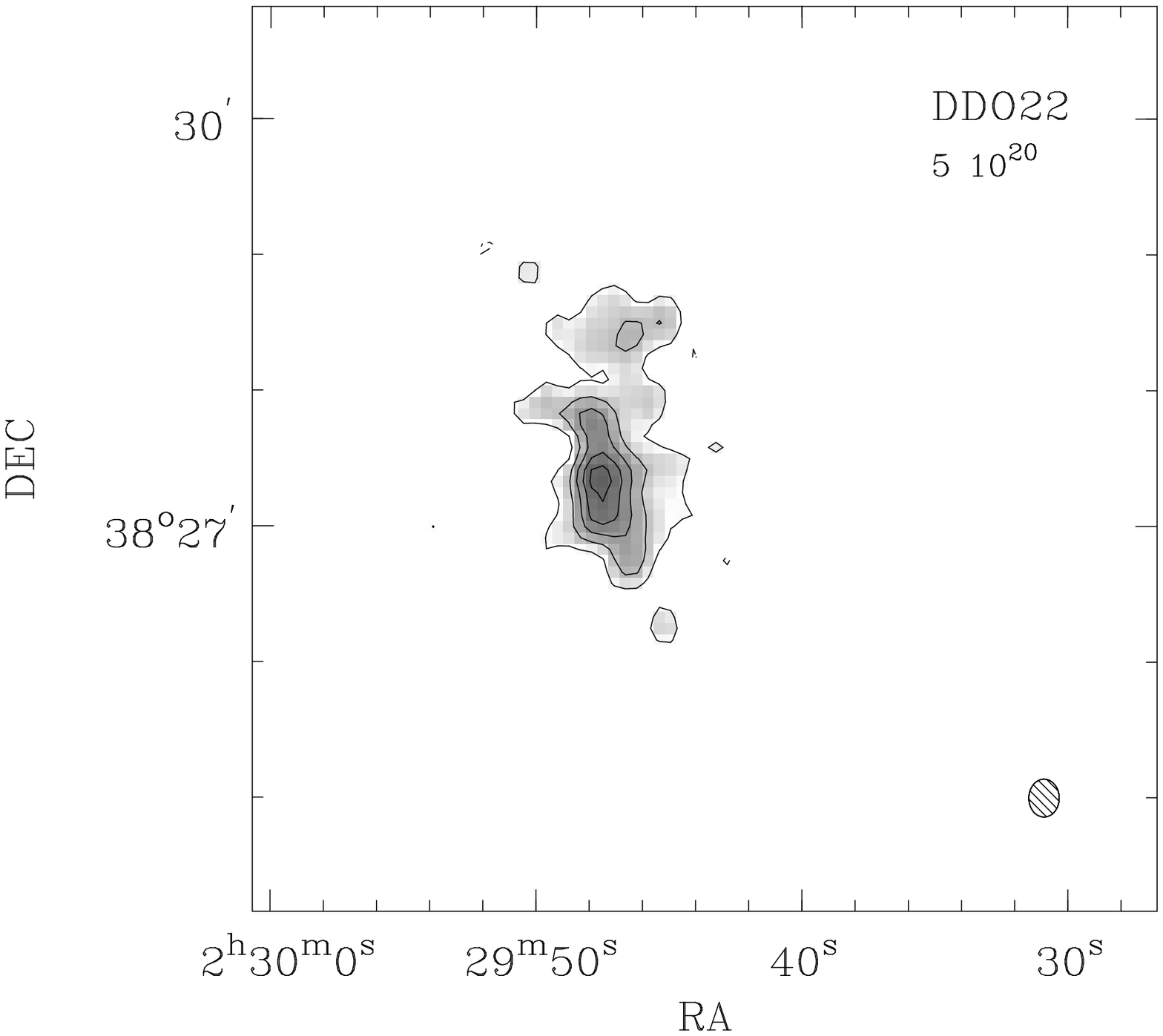}}
\end{minipage}
\hfill
\begin{minipage}[b]{5.7cm}
\resizebox{5.7cm}{!}{\includegraphics[angle=-90,clip=true]{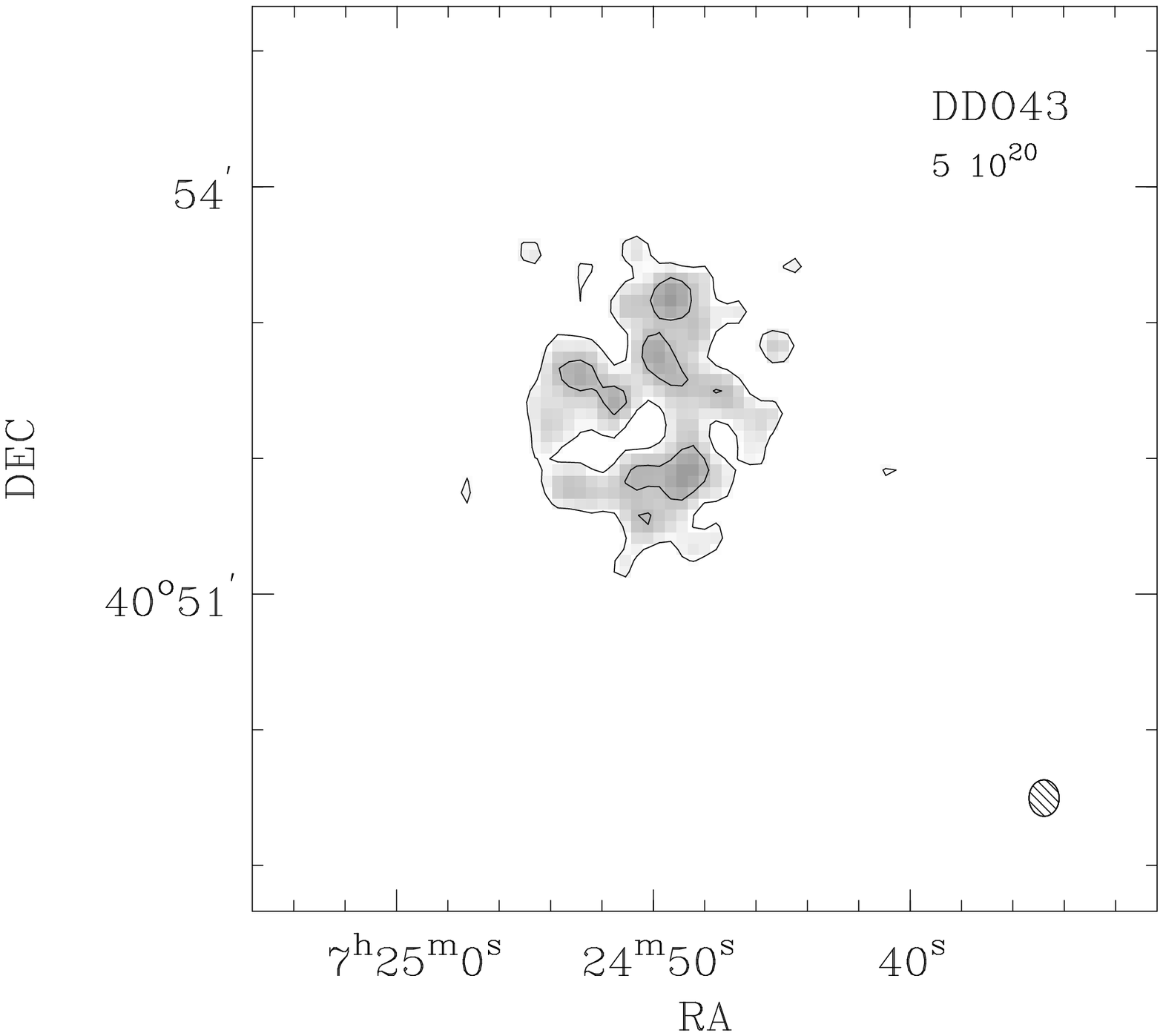}}
\end{minipage}
\hfill
\begin{minipage}[b]{5.7cm}
\resizebox{5.7cm}{!}{\includegraphics[angle=-90,clip=true]{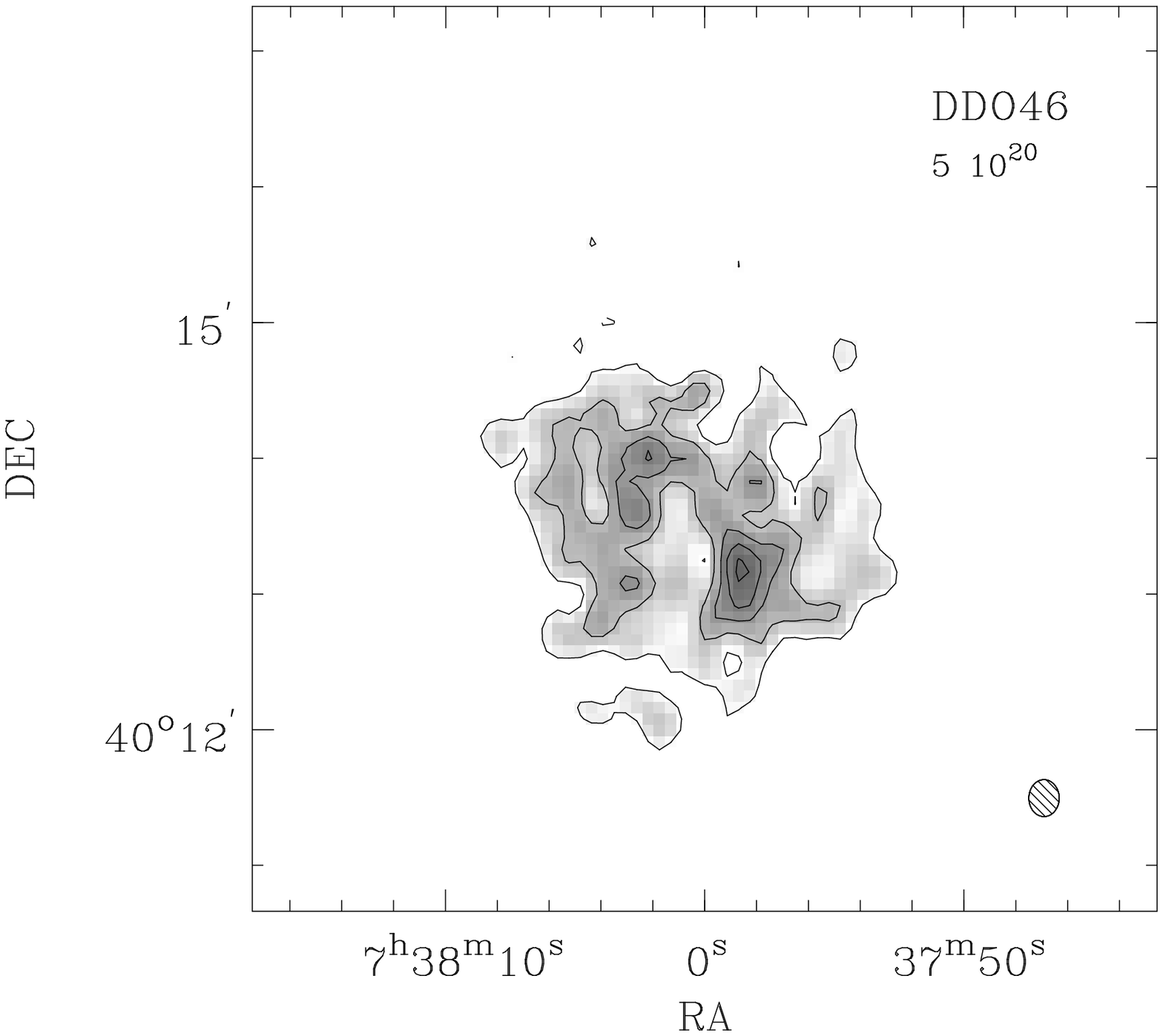}}
\end{minipage}
\begin{minipage}[t]{5.7cm}
\resizebox{5.7cm}{!}{\includegraphics[angle=-90,clip=true]{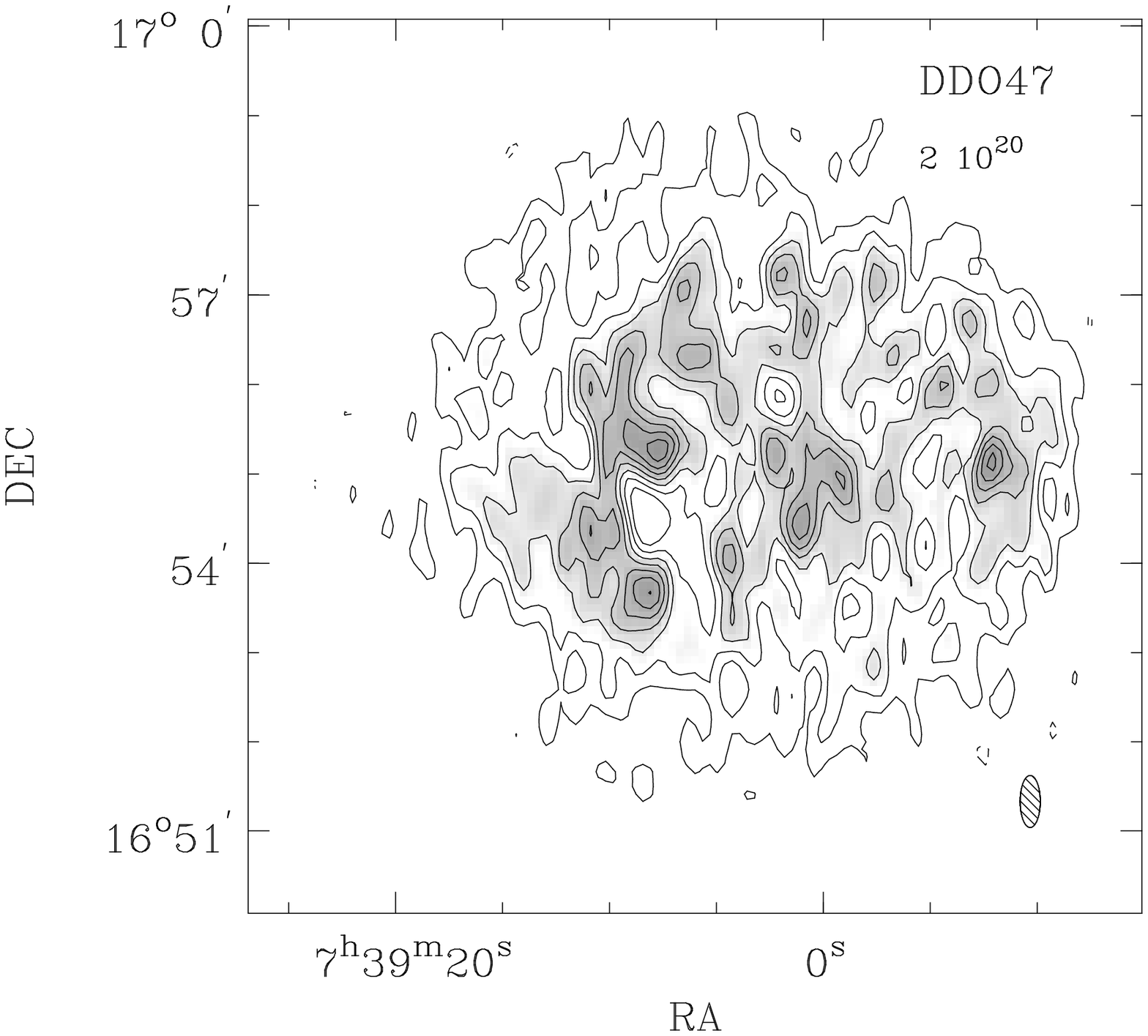}}
\noindent
\end{minipage}
\hfill
\begin{minipage}[b]{5.7cm}
\resizebox{5.7cm}{!}{\includegraphics[angle=-90,clip=true]{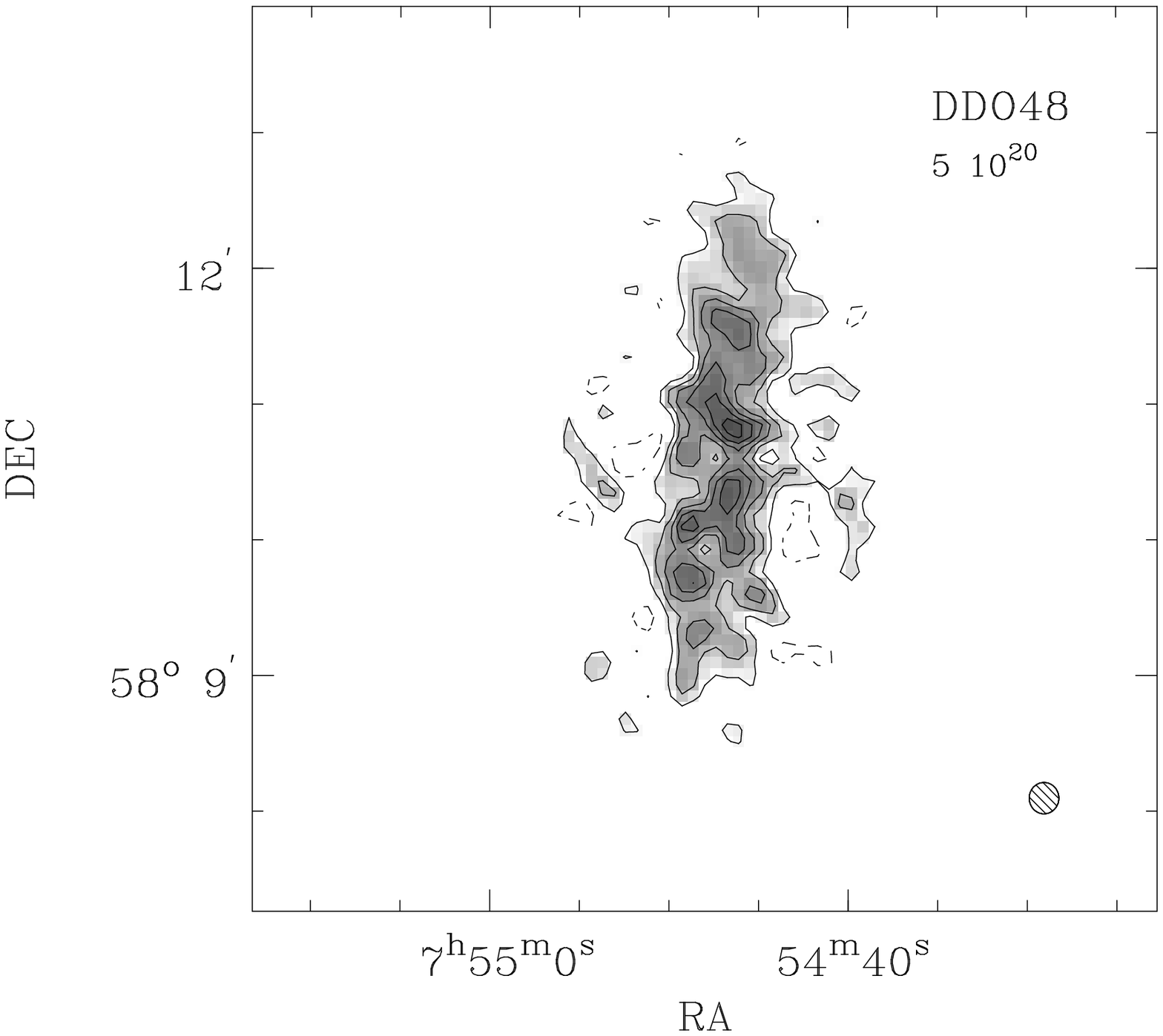}}
\noindent
\end{minipage}
\hfill
\begin{minipage}[b]{5.7cm}
\resizebox{5.7cm}{!}{\includegraphics[angle=-90,clip=true]{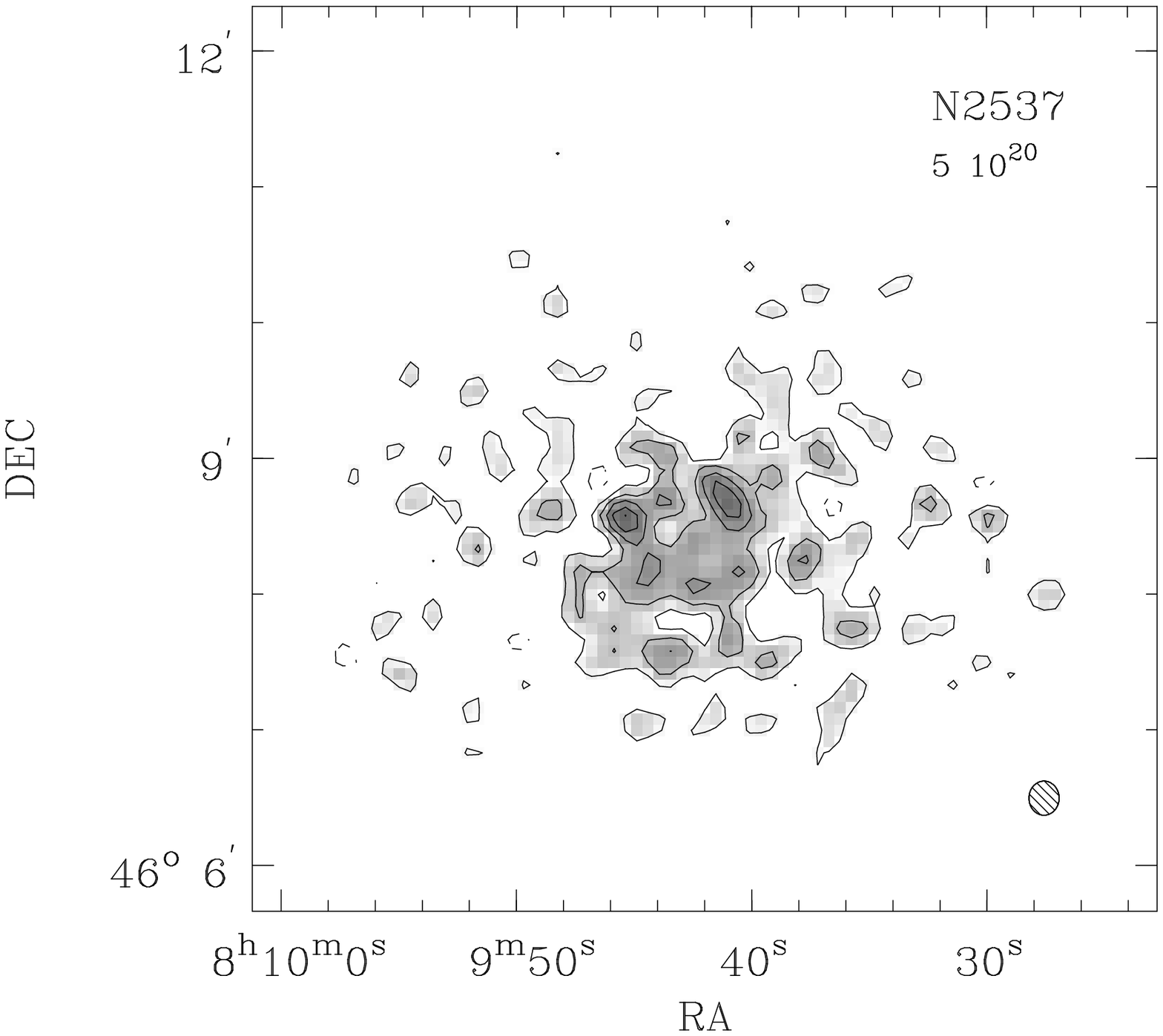}}
\end{minipage} 
\begin{minipage}[t]{5.7cm}
\resizebox{5.7cm}{!}{\includegraphics[angle=-90,clip=true]{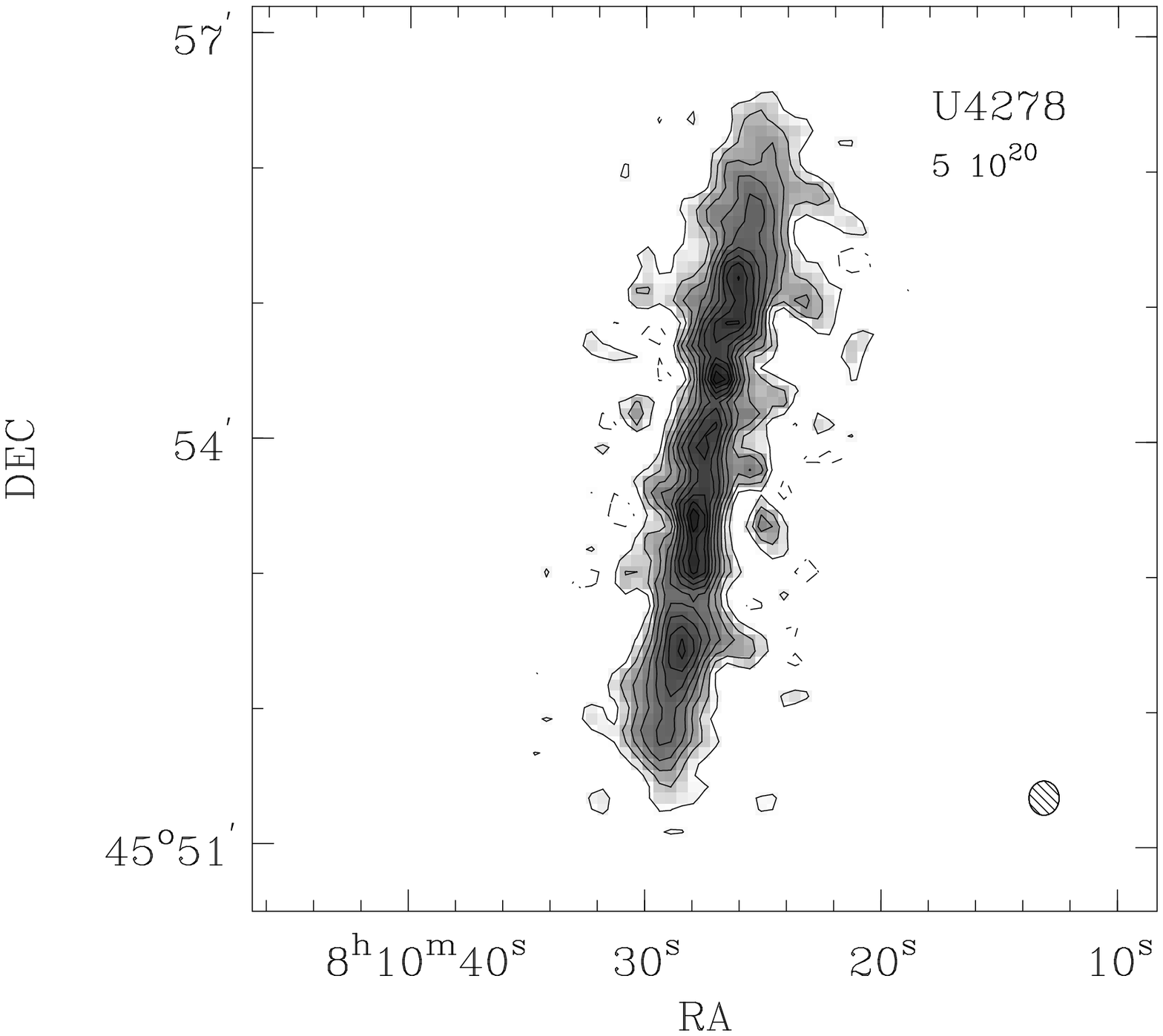}}
\end{minipage} 
\hfill
\begin{minipage}[b]{5.7cm}
\resizebox{5.7cm}{!}{\includegraphics[angle=-90,clip=true]{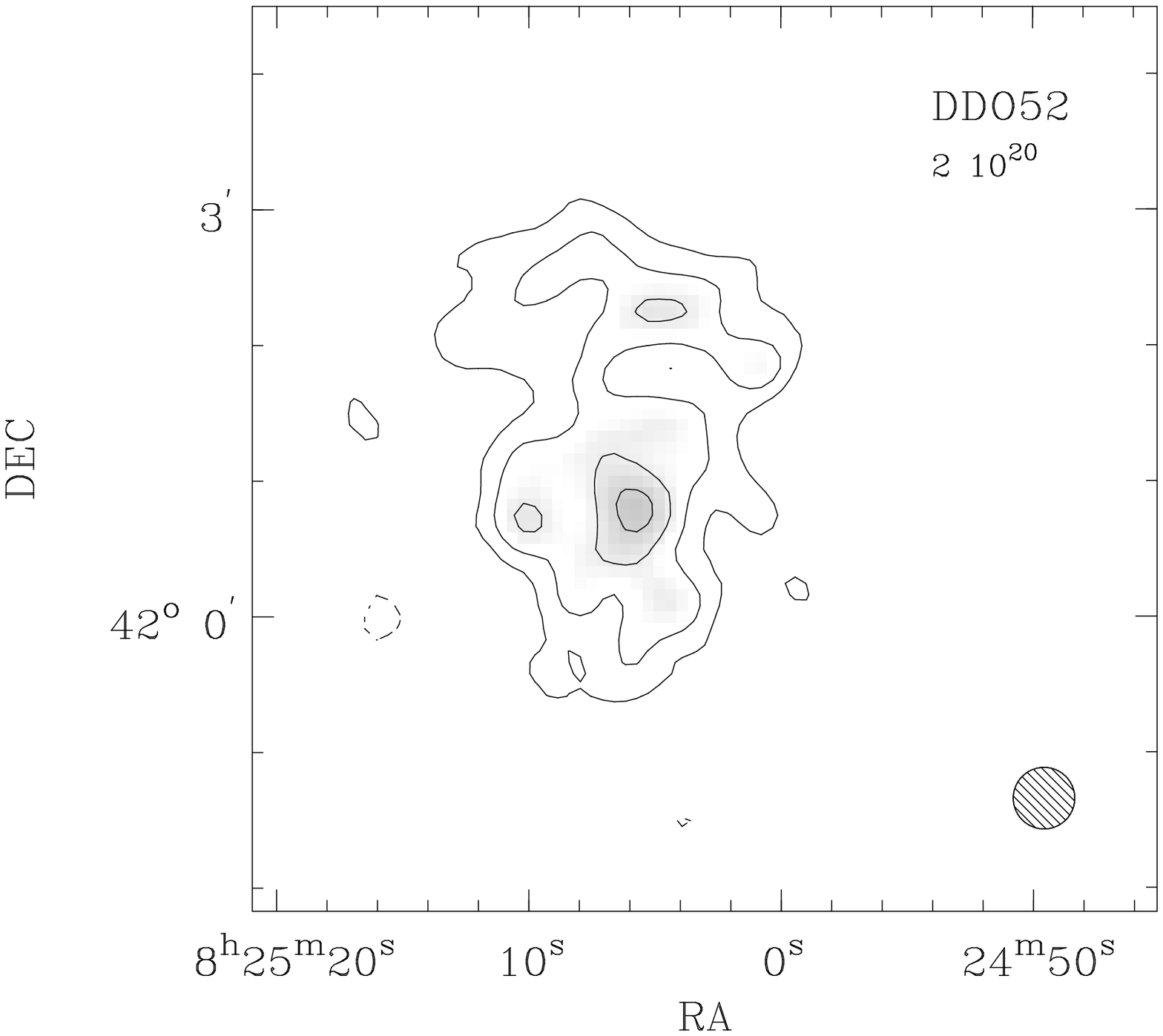}}
\noindent
\end{minipage}
\hfill
\begin{minipage}[b]{5.7cm}
\resizebox{5.7cm}{!}{\includegraphics[angle=-90,clip=true]{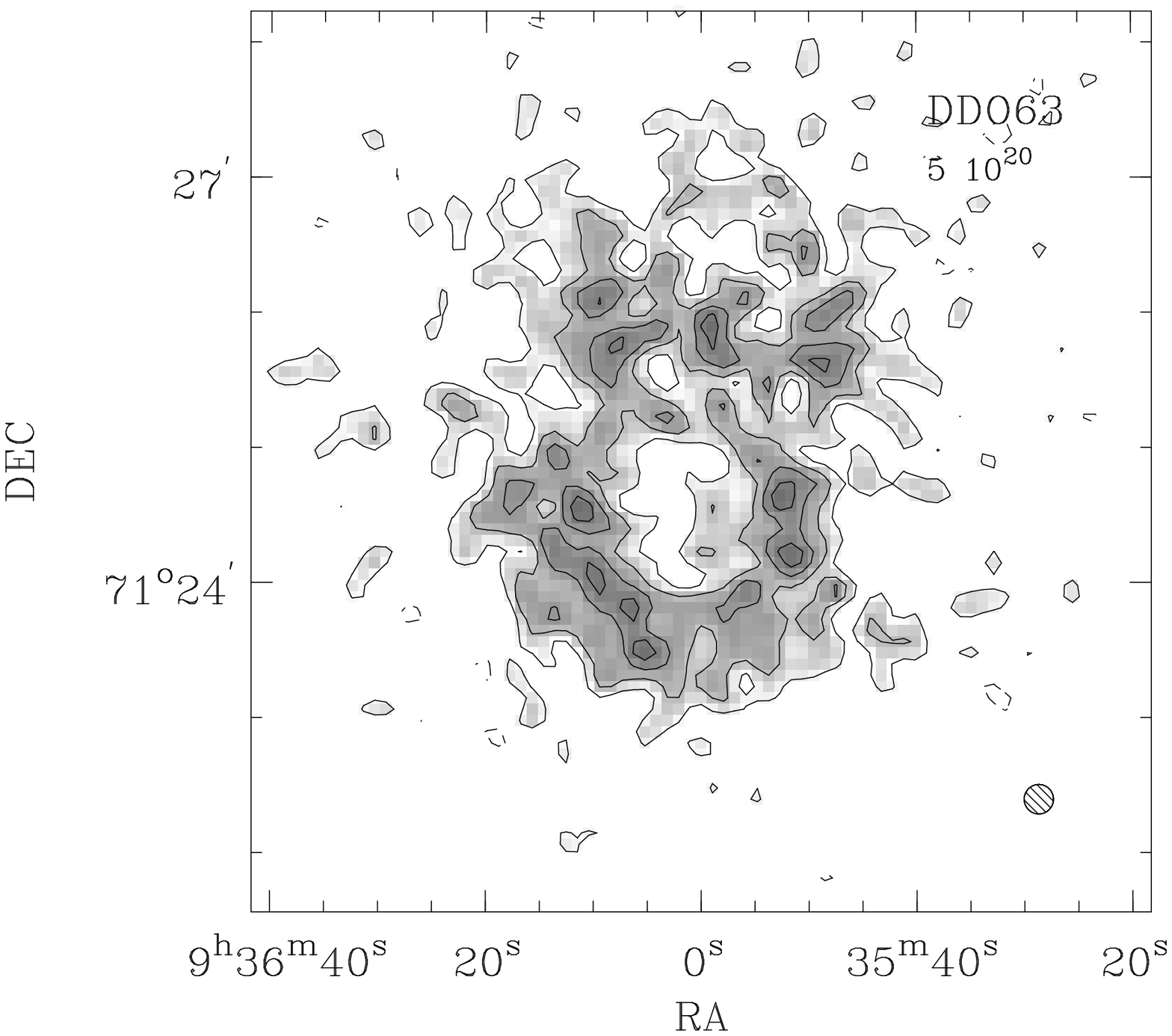}}
\noindent
\end{minipage}
\begin{minipage}[t]{5.7cm}
\resizebox{5.7cm}{!}{\includegraphics[angle=-90,clip=true]{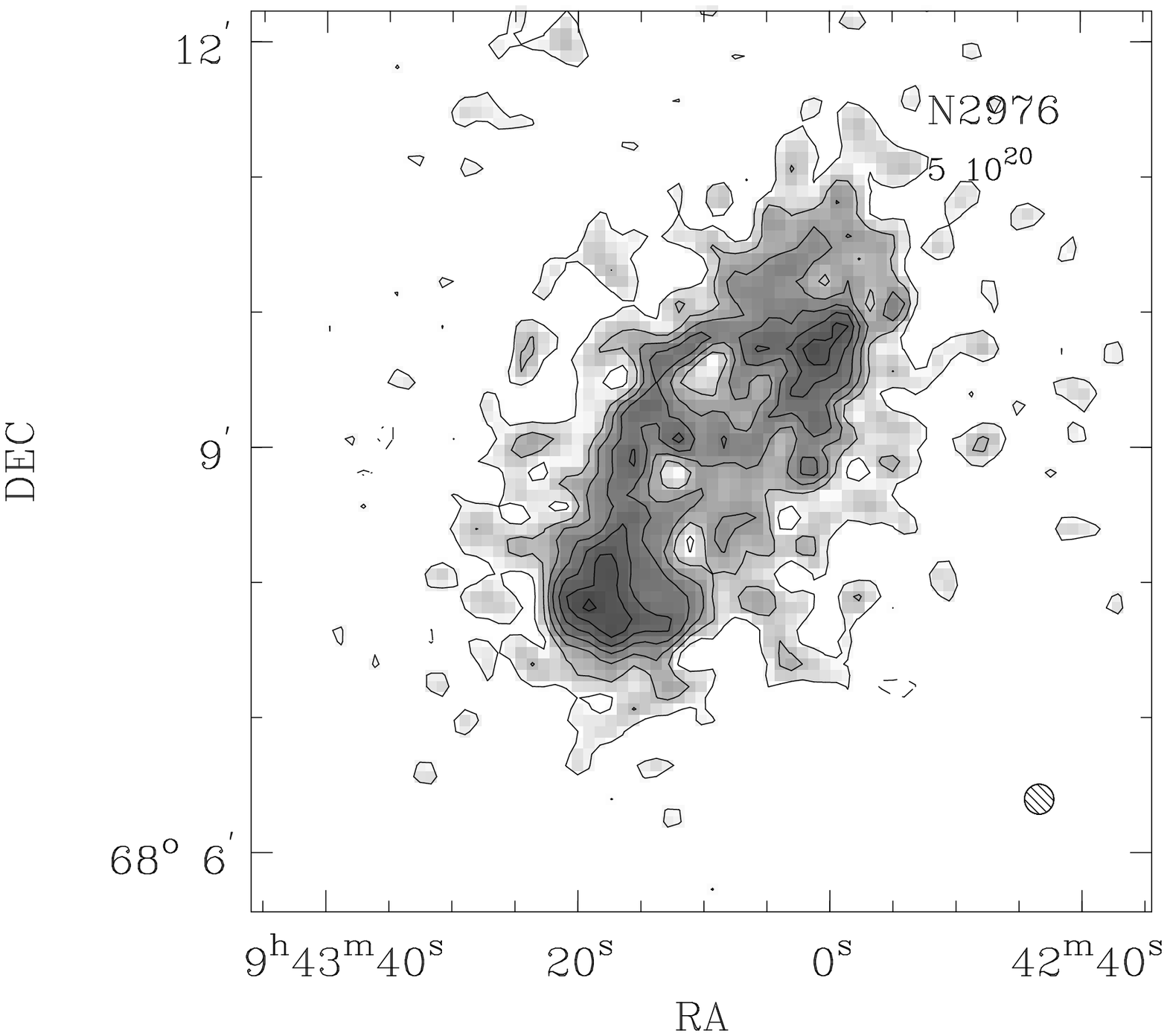}}
\noindent
\end{minipage}
\hfill
\begin{minipage}[b]{5.7cm}
\resizebox{5.7cm}{!}{\includegraphics[angle=-90,clip=true]{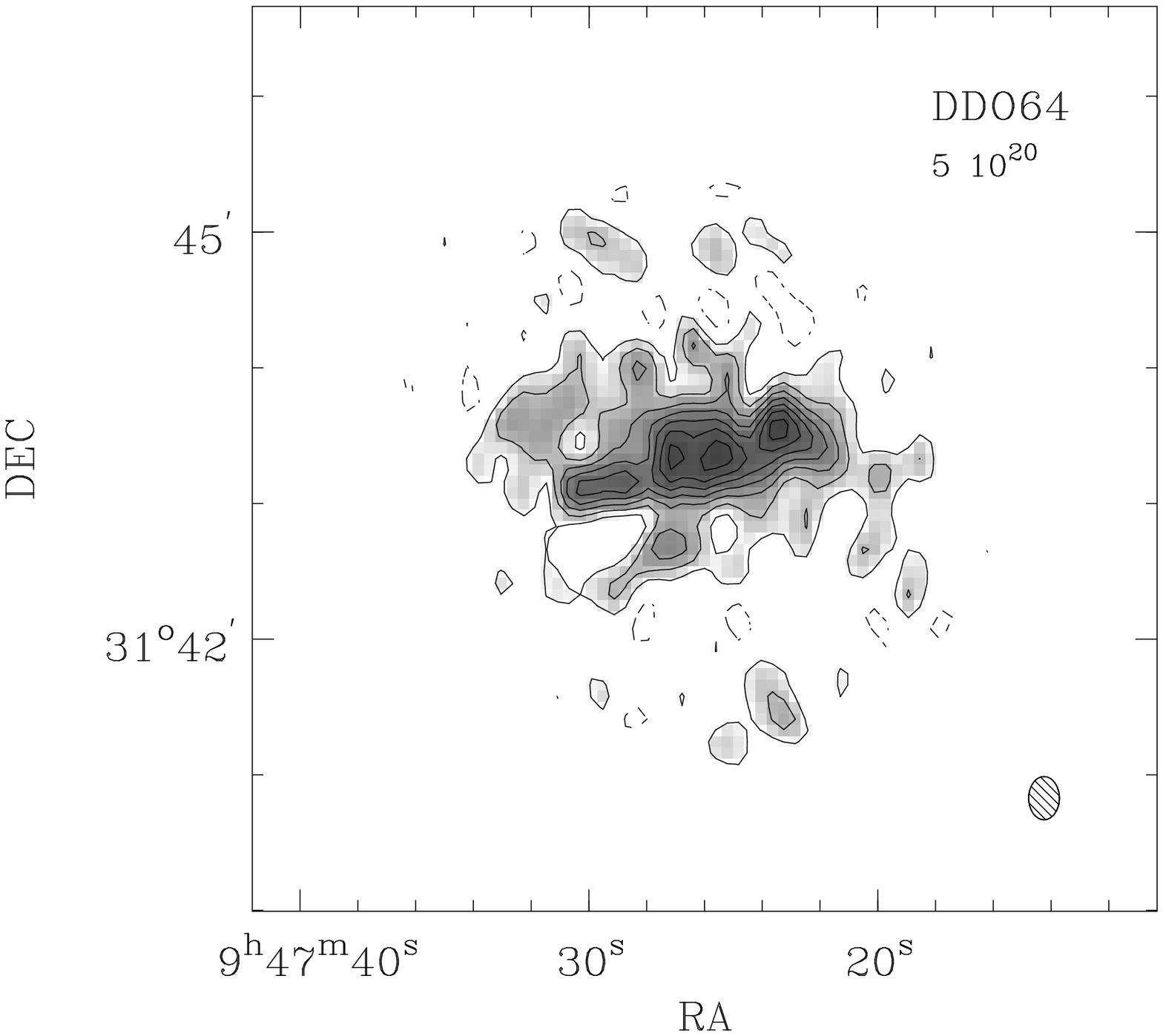}}
\noindent
\end{minipage}
\hfill
\begin{minipage}[b]{5.7cm}
\resizebox{5.7cm}{!}{\includegraphics[angle=-90,clip=true]{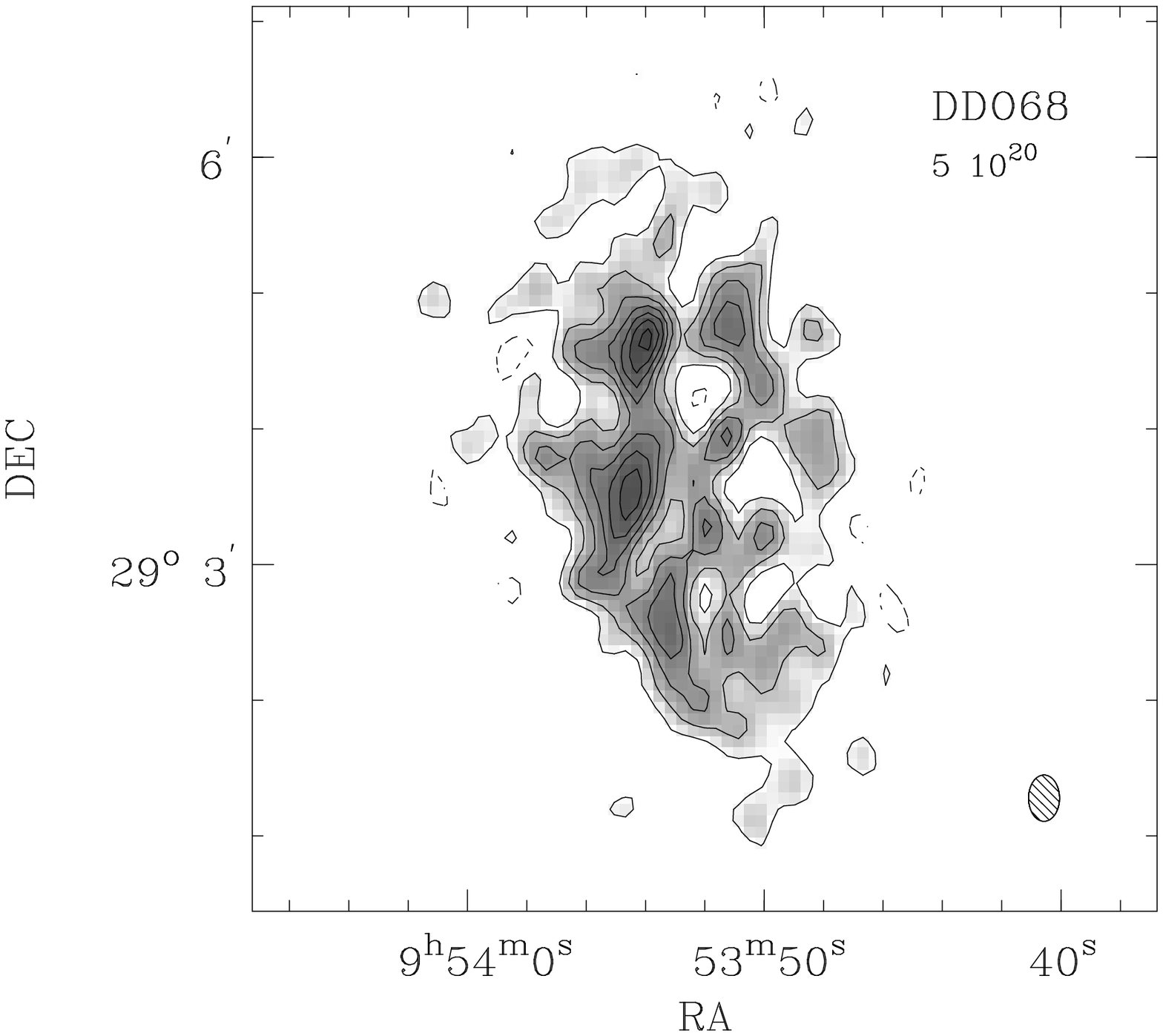}}
\noindent
\end{minipage}
\caption{HI column density maps at 13.5$''$ resolution. 
Contours ($\cm2$) are in steps printed below object name. 
%
}
\end{figure*}

\begin{figure*}
\addtocounter{figure}{-1}
\begin{minipage}[t]{5.7cm}
\resizebox{5.7cm}{!}{\includegraphics[angle=-90,clip=true]{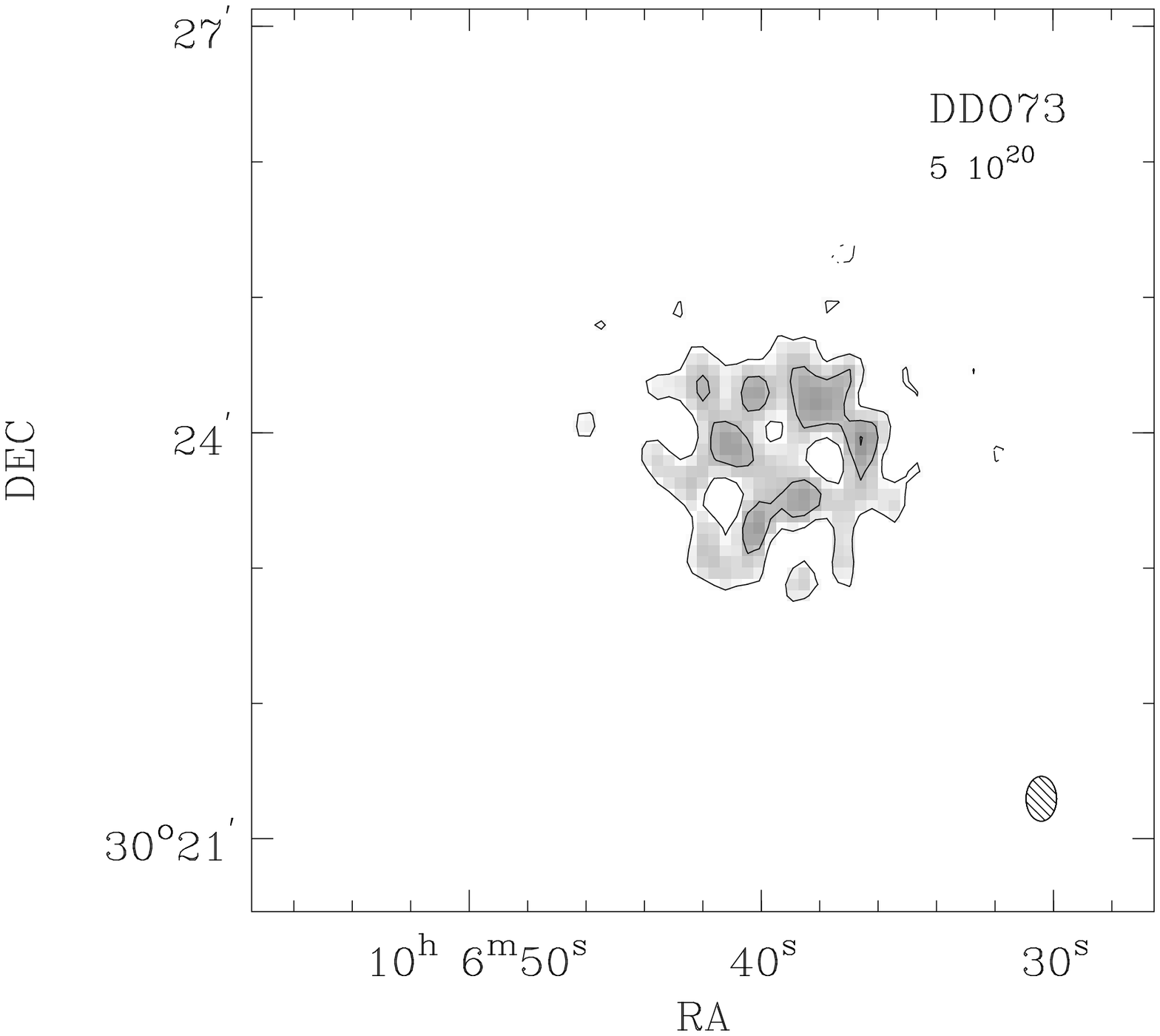}}
\noindent
\end{minipage}
\hfill
\begin{minipage}[b]{5.7cm}
\resizebox{5.7cm}{!}{\includegraphics[angle=-90,clip=true]{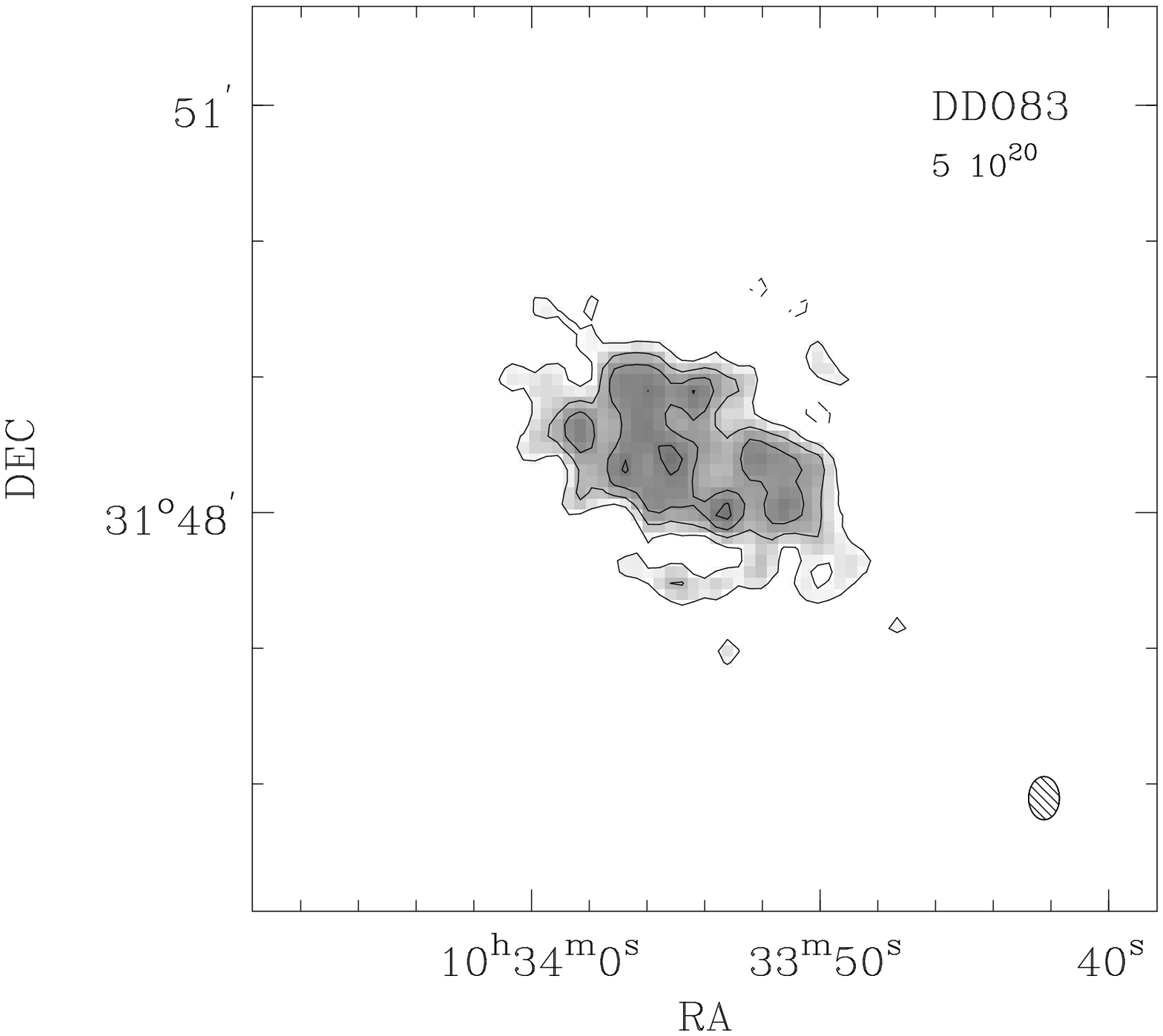}}
\noindent
\end{minipage}
\hfill
\begin{minipage}[b]{5.7cm}
\resizebox{5.7cm}{!}{\includegraphics[angle=-90,clip=true]{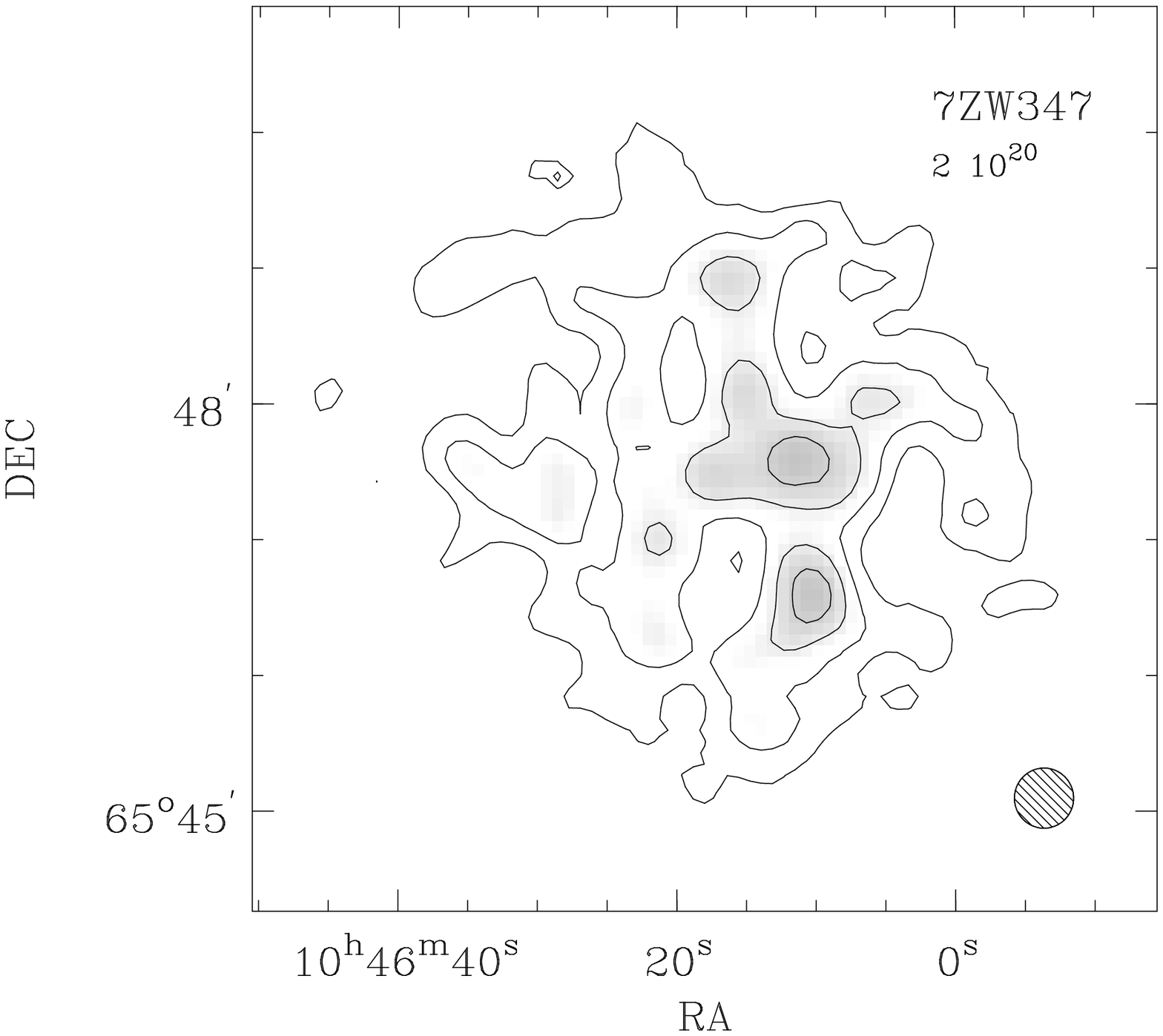}}
\noindent
\end{minipage}
\begin{minipage}[t]{5.7cm}
\resizebox{5.7cm}{!}{\includegraphics[angle=-90,clip=true]{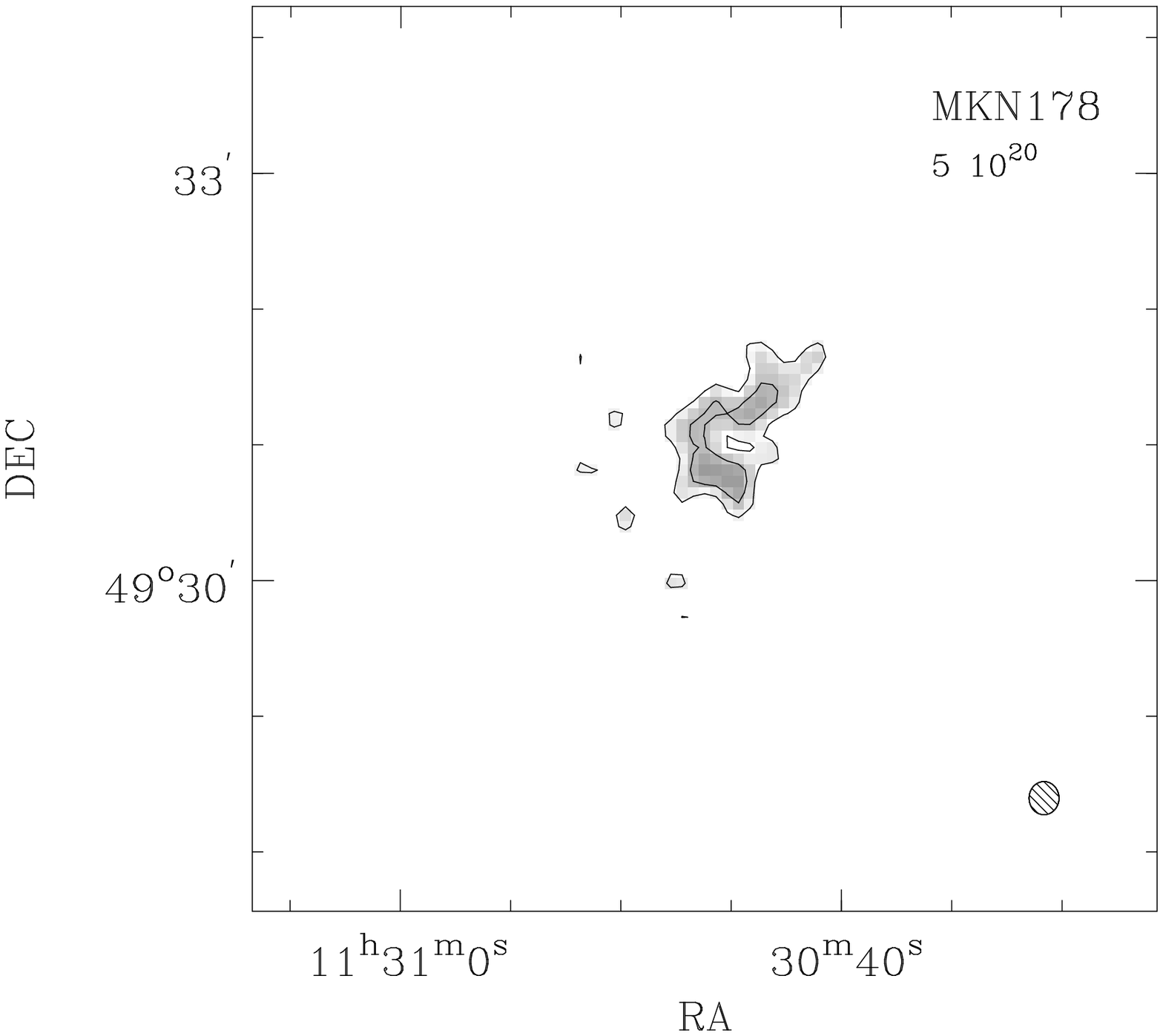}}
\noindent
\end{minipage}
\hfill
\begin{minipage}[b]{5.7cm}
\resizebox{5.7cm}{!}{\includegraphics[angle=-90,clip=true]{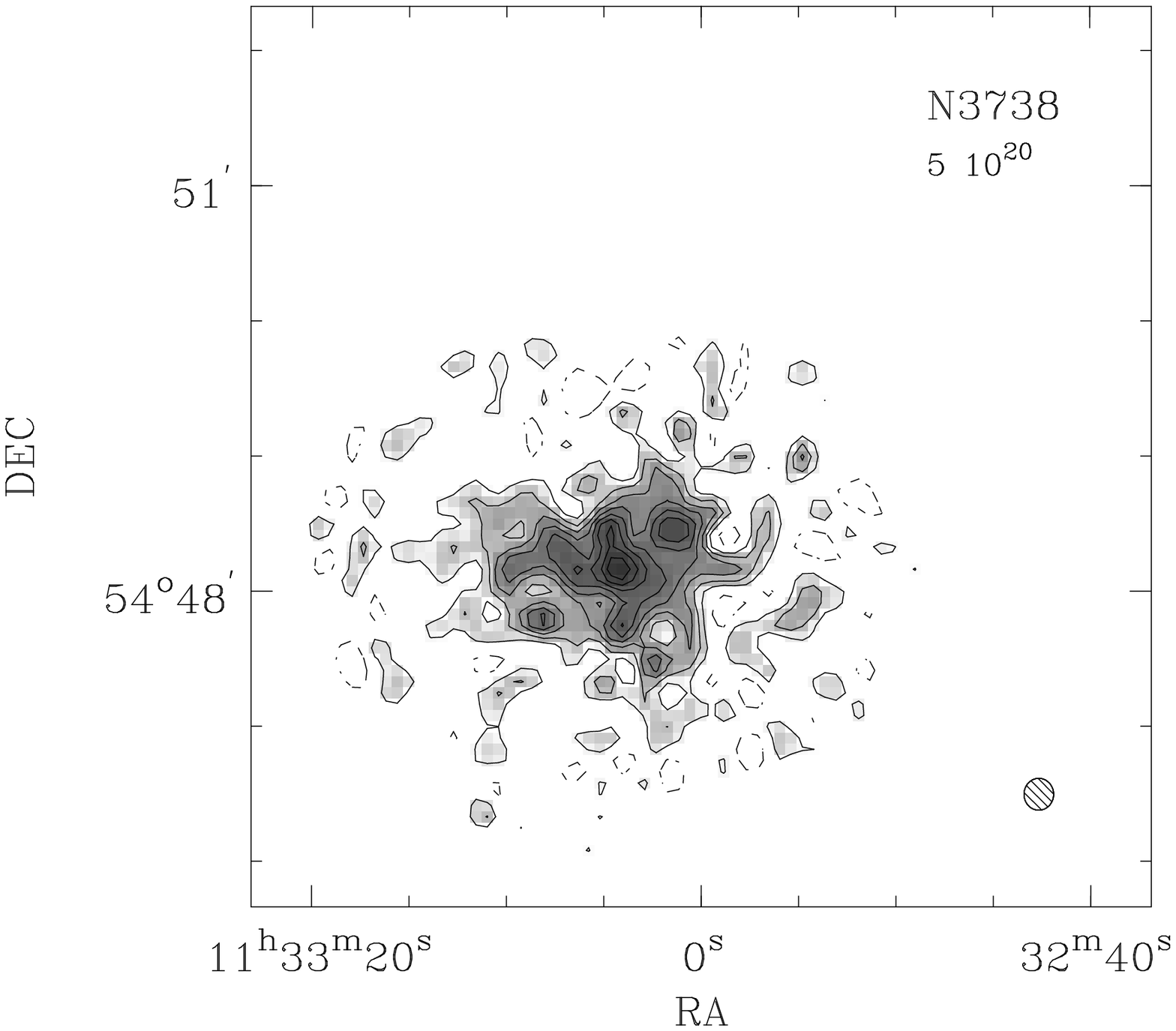}}
\noindent
\end{minipage}
\hfill
\begin{minipage}[b]{5.7cm}
\resizebox{5.7cm}{!}{\includegraphics[angle=-90,clip=true]{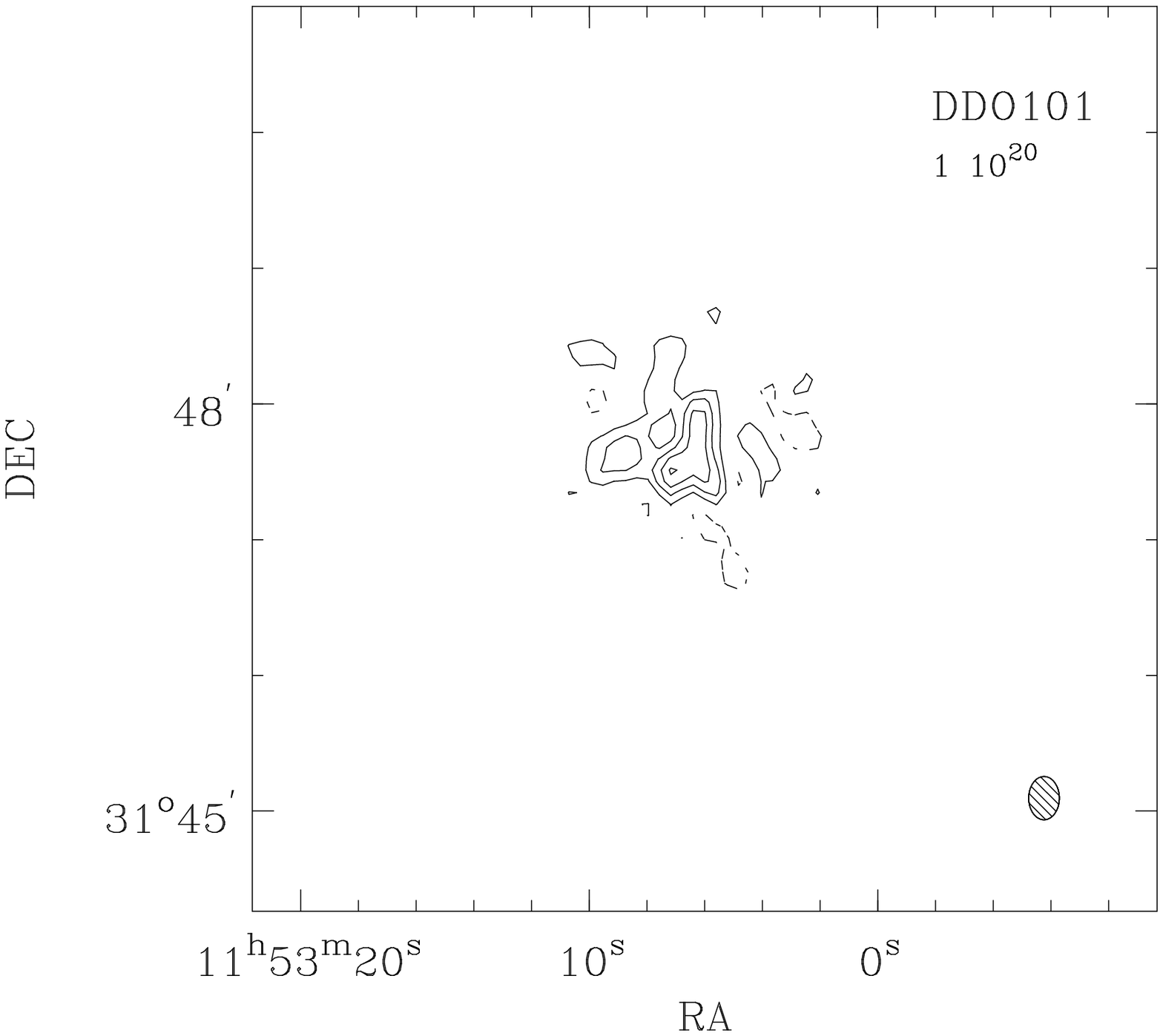}}
\noindent
\end{minipage}
\begin{minipage}[t]{5.7cm}
\resizebox{5.7cm}{!}{\includegraphics[angle=-90,clip=true]{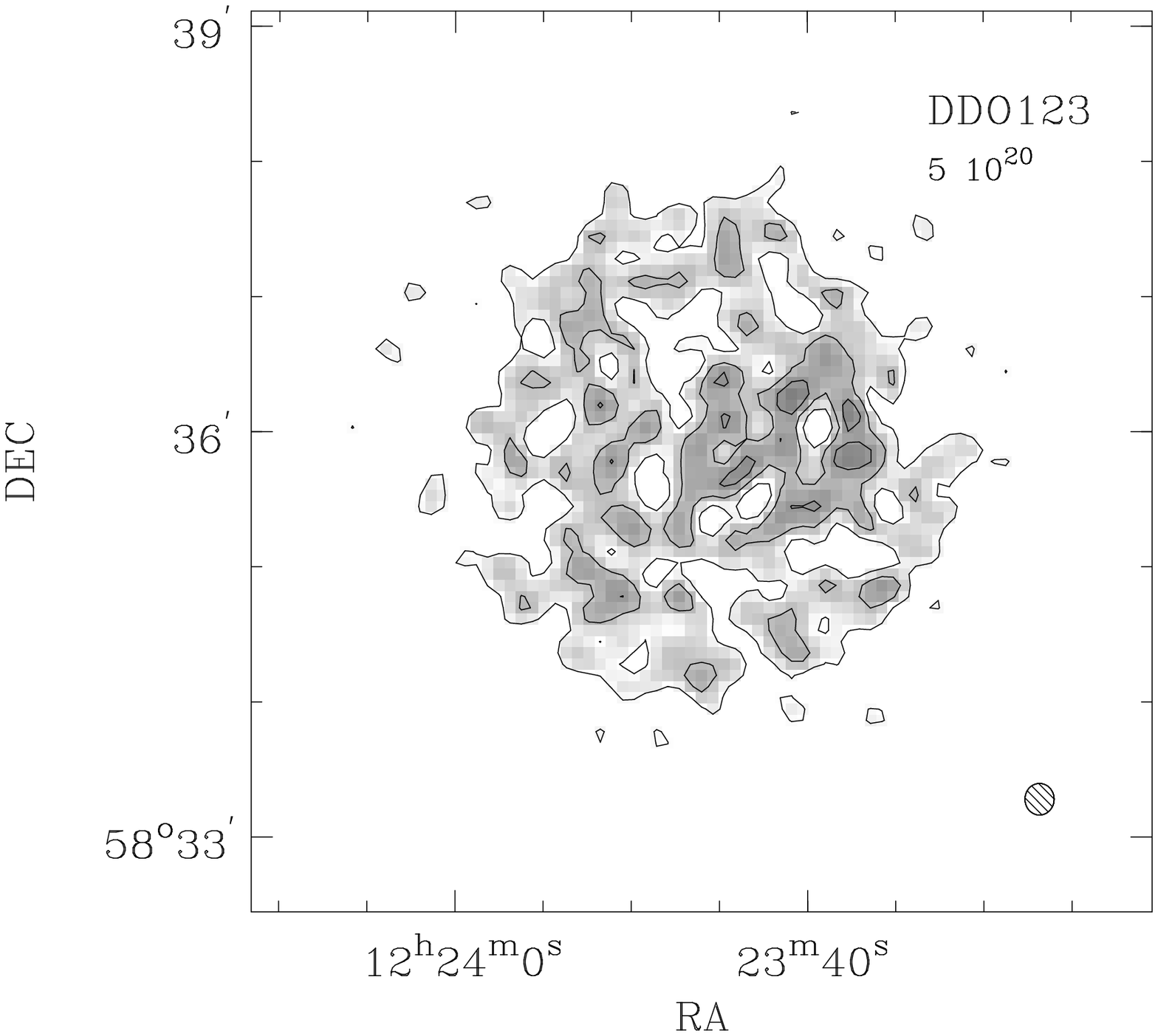}}
\noindent
\end{minipage}
\hfill
\begin{minipage}[b]{5.7cm}
\resizebox{5.7cm}{!}{\includegraphics[angle=-90,clip=true]{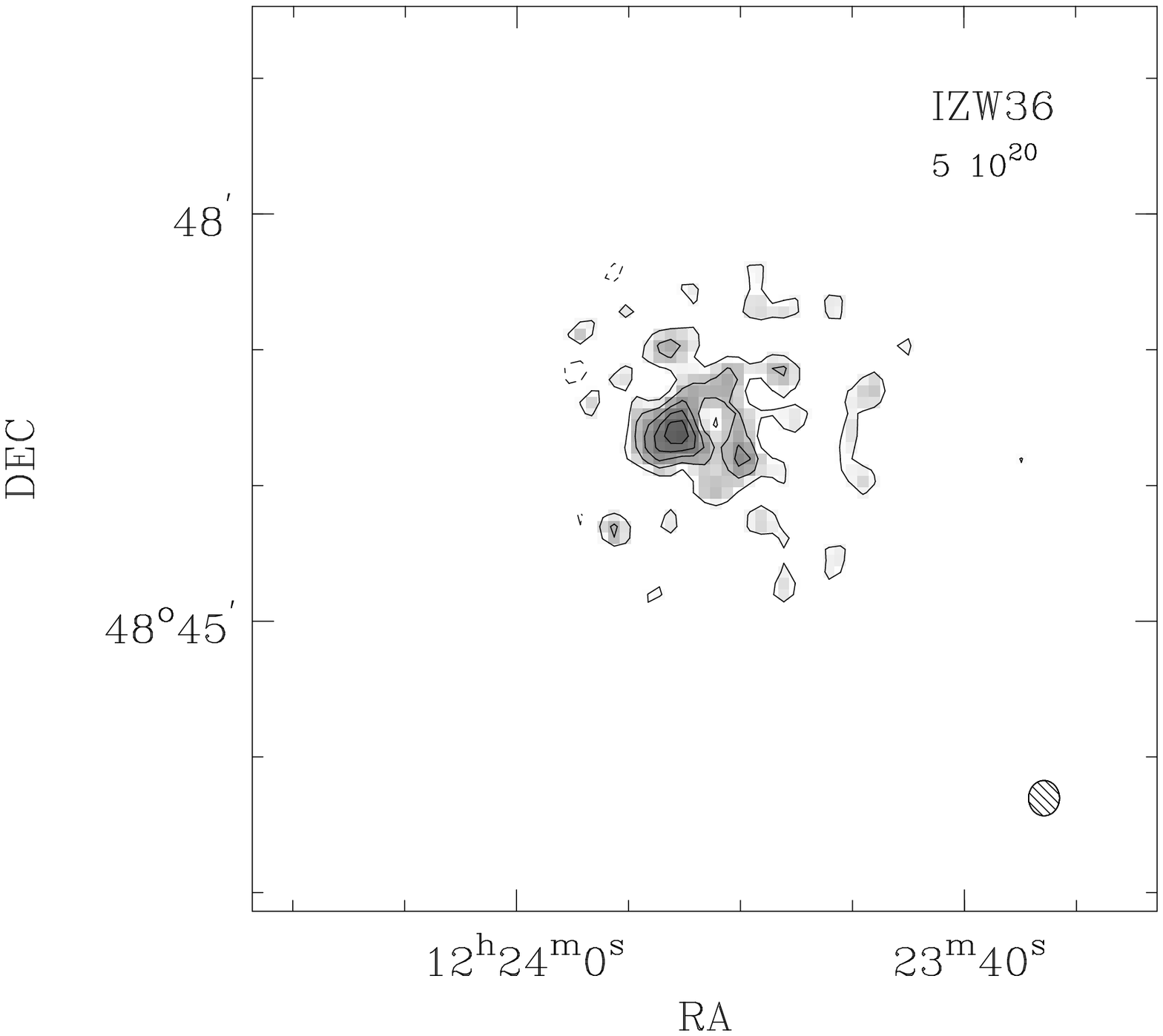}}
\noindent
\end{minipage}
\hfill
\begin{minipage}[b]{5.7cm}
\resizebox{5.7cm}{!}{\includegraphics[angle=-90,clip=true]{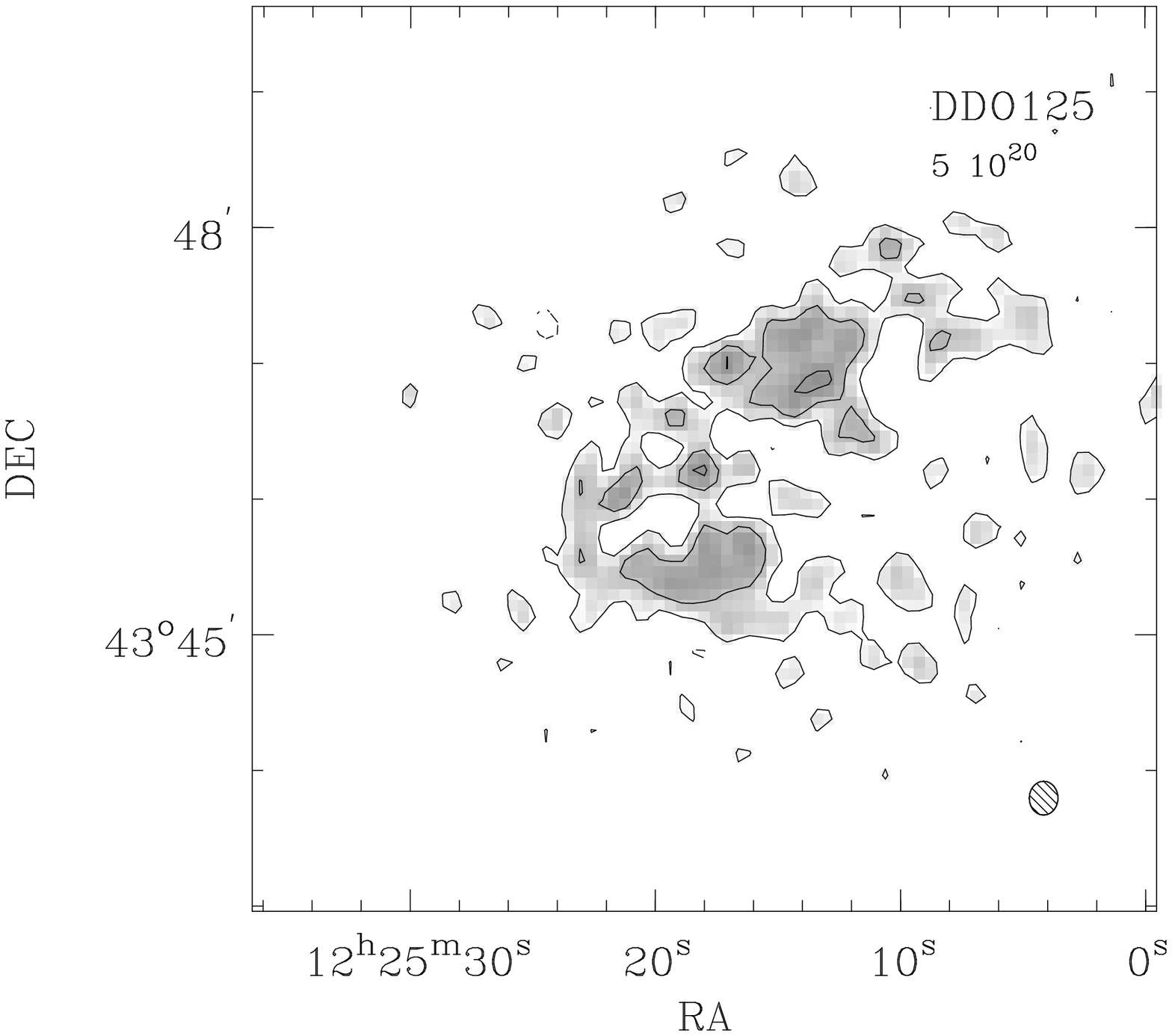}}
\noindent
\end{minipage}
\begin{minipage}[t]{5.5cm}
\resizebox{5.5cm}{!}{\includegraphics[angle=-90,clip=true]{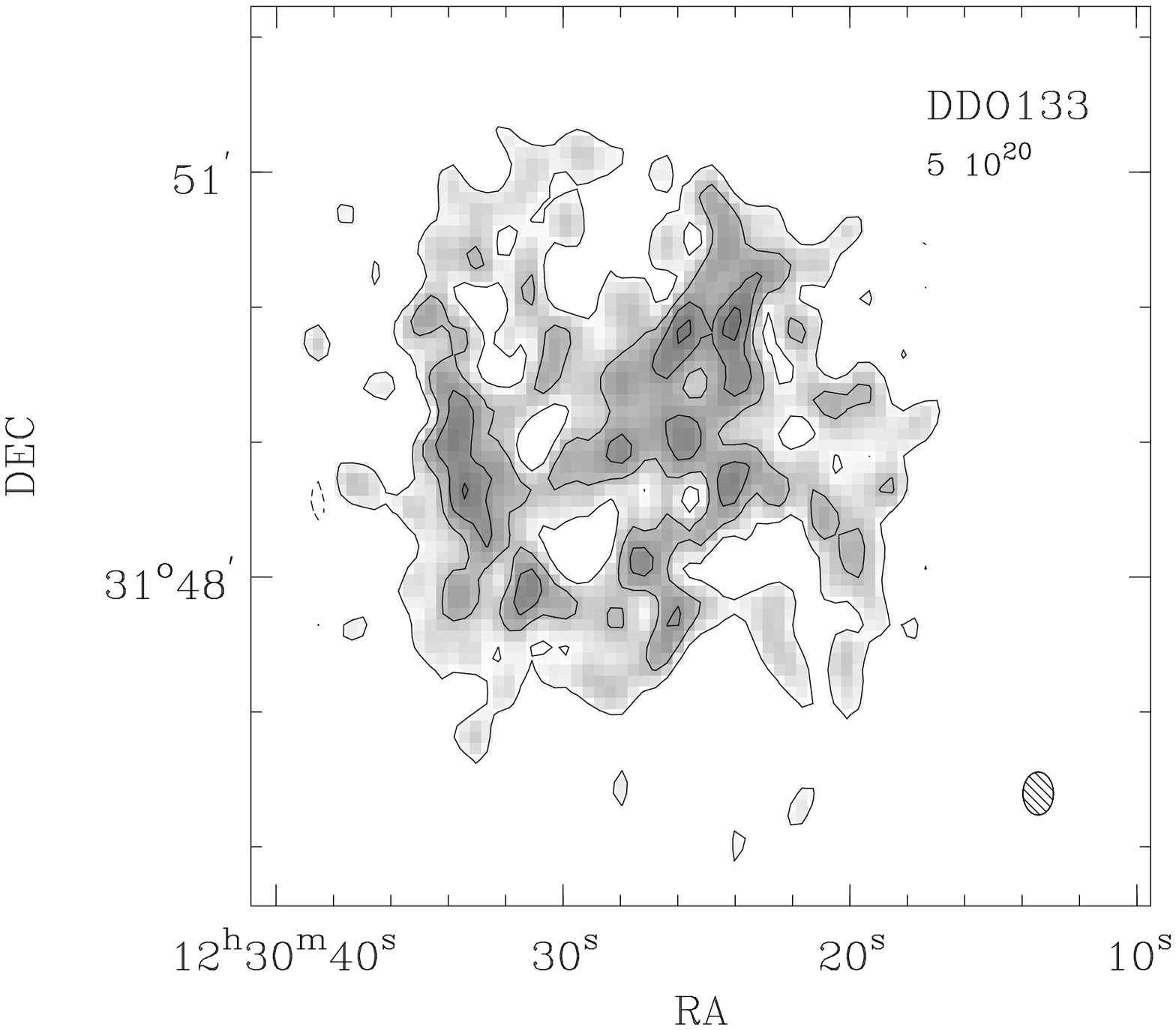}}
\end{minipage}
\hfill
\begin{minipage}[b]{5.7cm}
\noindent
\resizebox{5.7cm}{!}{\includegraphics[angle=-90,clip=true]{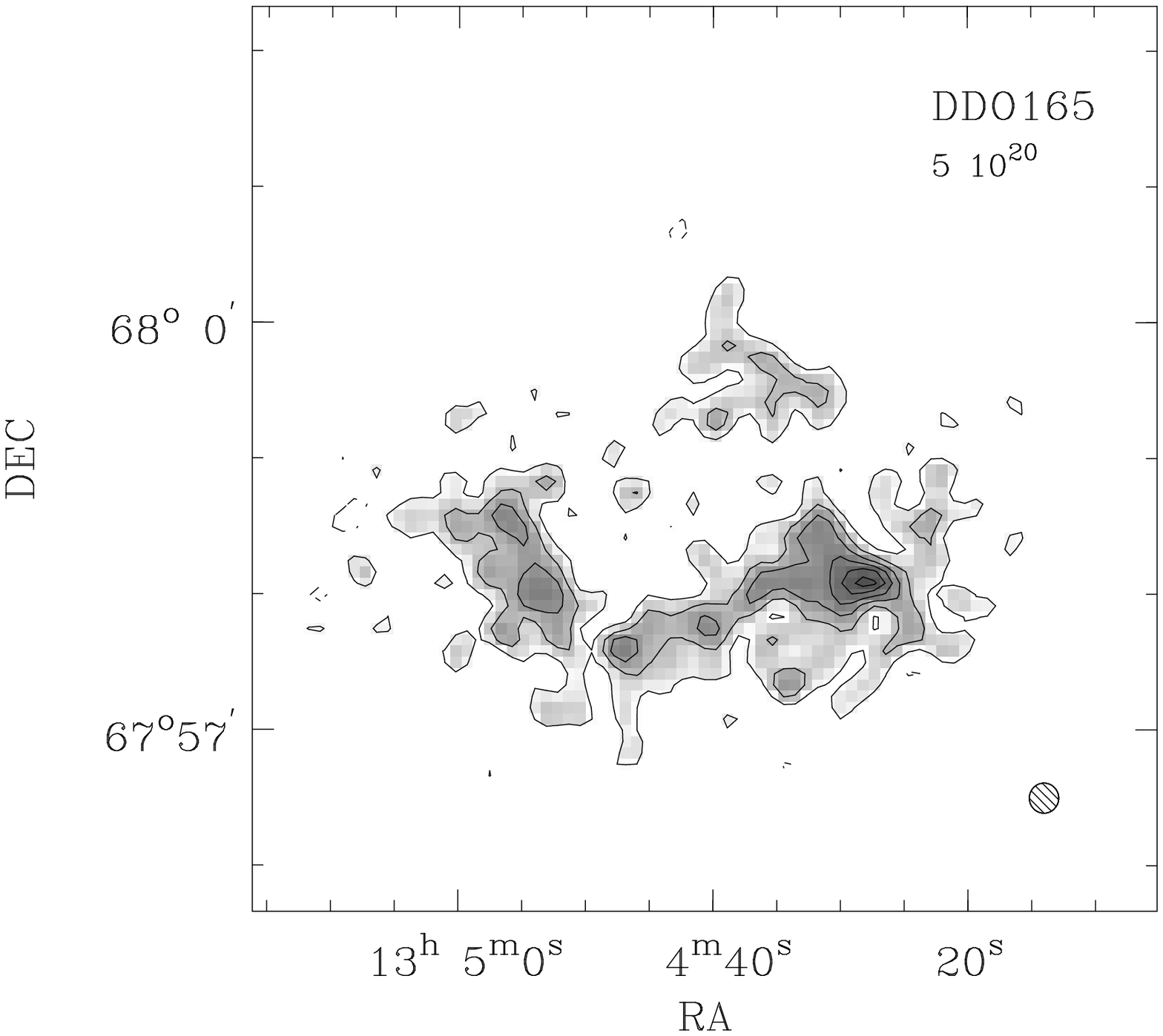}}
\end{minipage}
\hfill
\begin{minipage}[b]{5.7cm}
\resizebox{5.7cm}{!}{\includegraphics[angle=-90,clip=true]{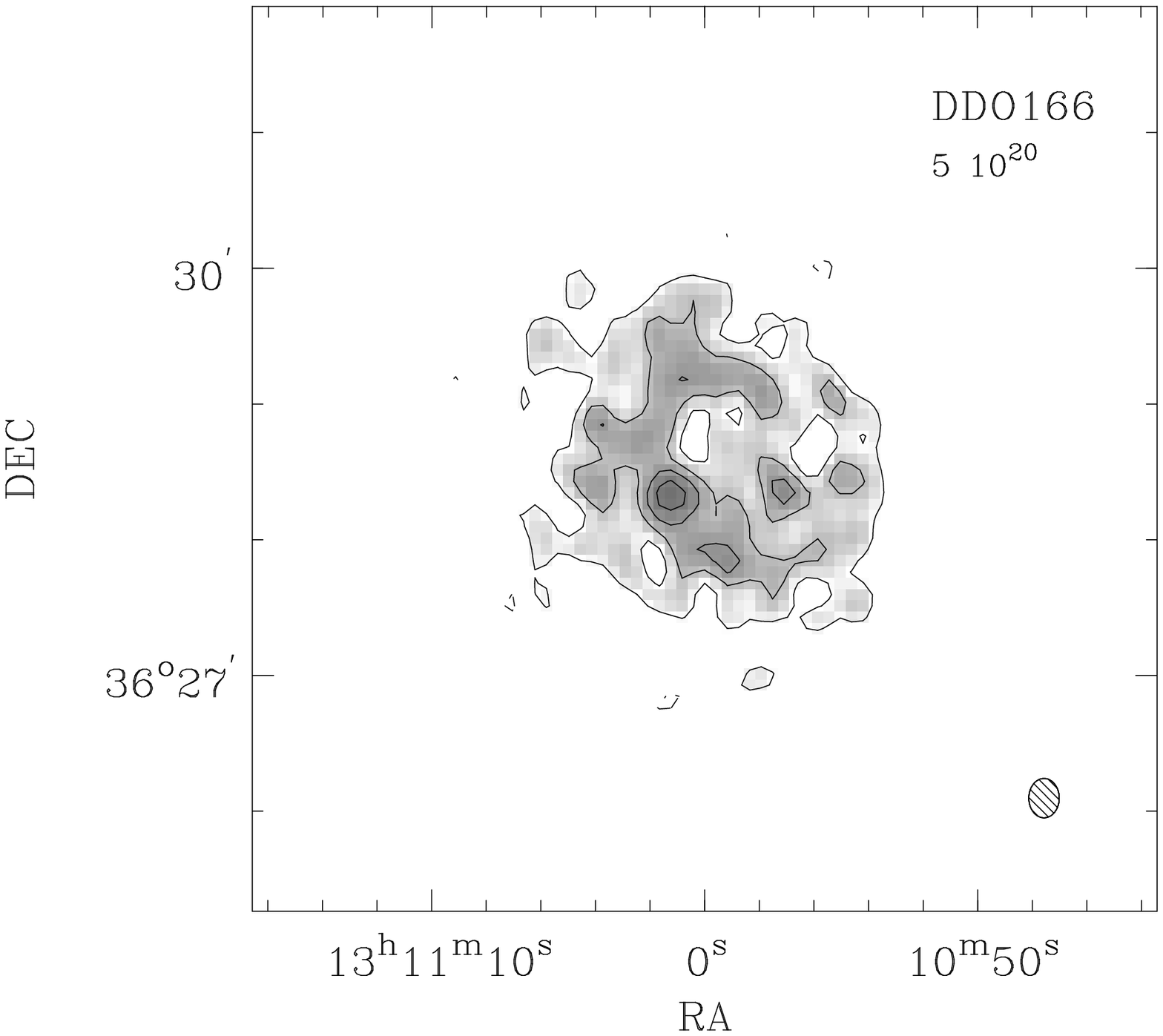}}
\noindent
\end{minipage}
\caption{continued; note: \small DDO\,52 and 7\small Zw\,347 maps are at 27$''$
resolution}
\end{figure*}

\begin{figure*}
\addtocounter{figure}{-1}
\begin{minipage}[t]{5.7cm}
\resizebox{5.7cm}{!}{\includegraphics[angle=-90,clip=true]{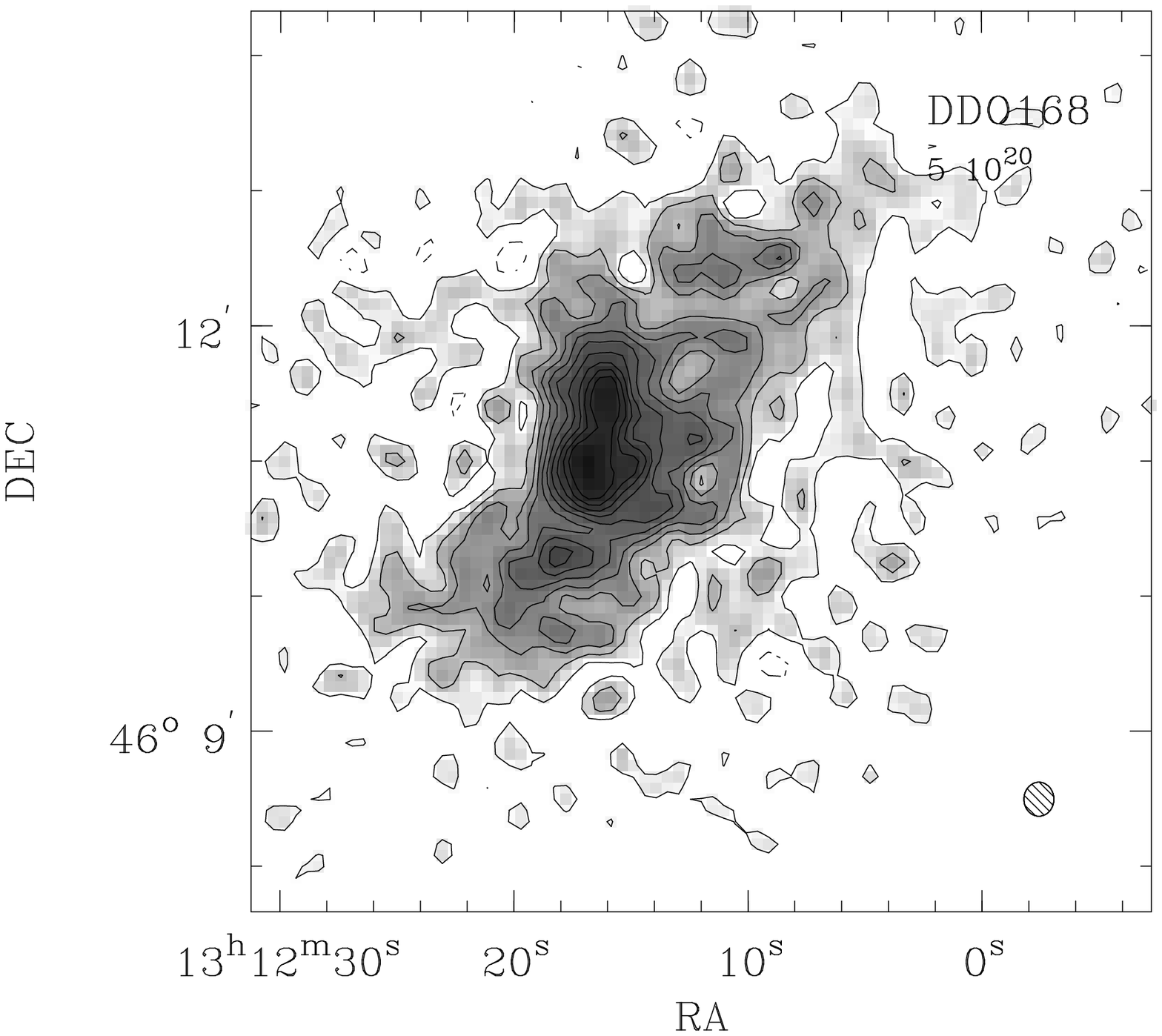}}
\end{minipage}
\hfill
\begin{minipage}[b]{5.7cm}
\resizebox{5.7cm}{!}{\includegraphics[angle=-90,clip=true]{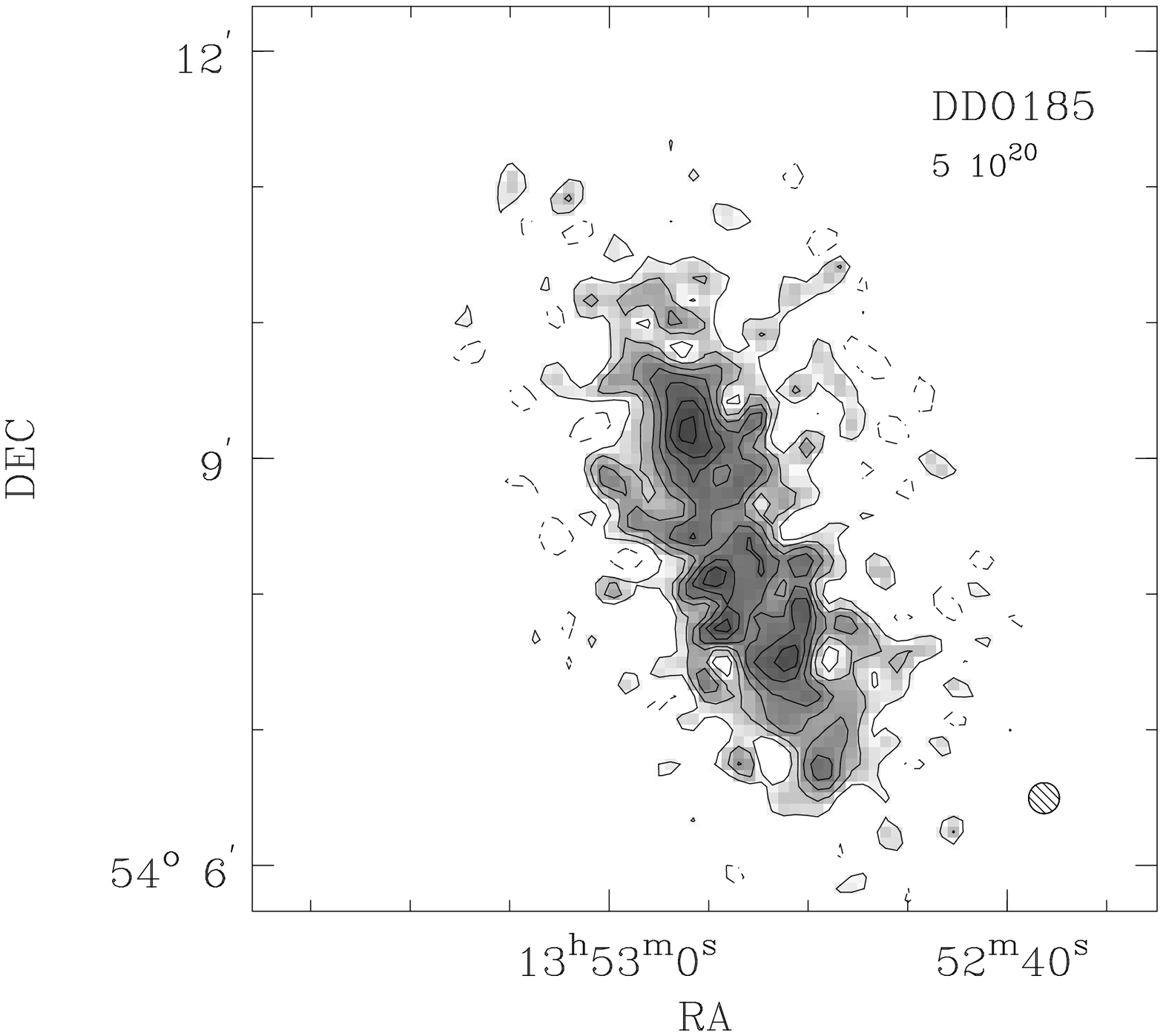}}
\end{minipage}
\hfill
\begin{minipage}[b]{5.7cm}
\resizebox{5.7cm}{!}{\includegraphics[angle=-90,clip=true]{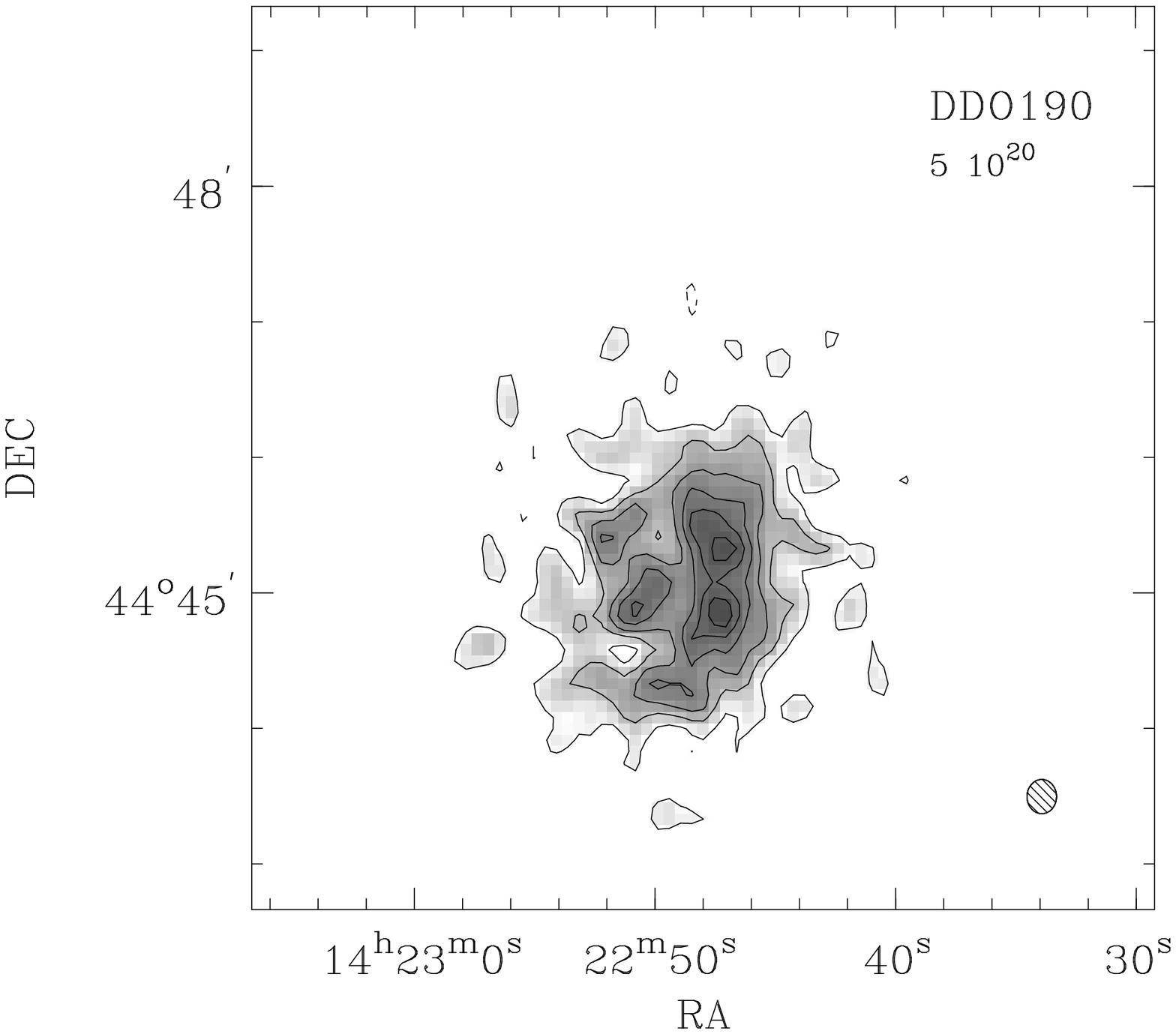}}
\noindent
\end{minipage}
\begin{minipage}[t]{5.7cm}
\resizebox{5.7cm}{!}{\includegraphics[angle=-90,clip=true]{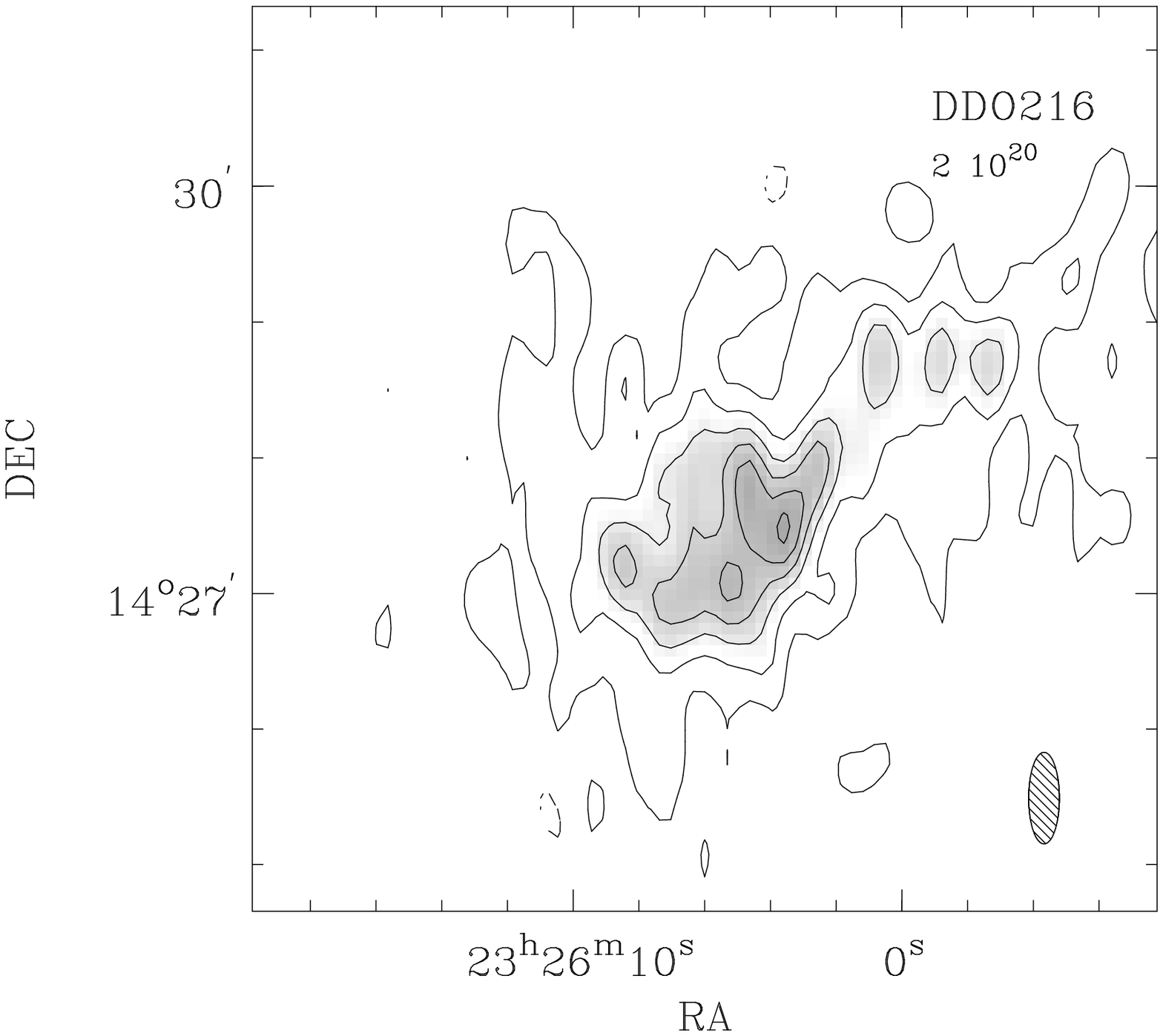}}
\noindent
\end{minipage}
\hspace{0.22cm}
\begin{minipage}[b]{5.7cm}
\resizebox{5.7cm}{!}{\includegraphics[angle=-90,clip=true]{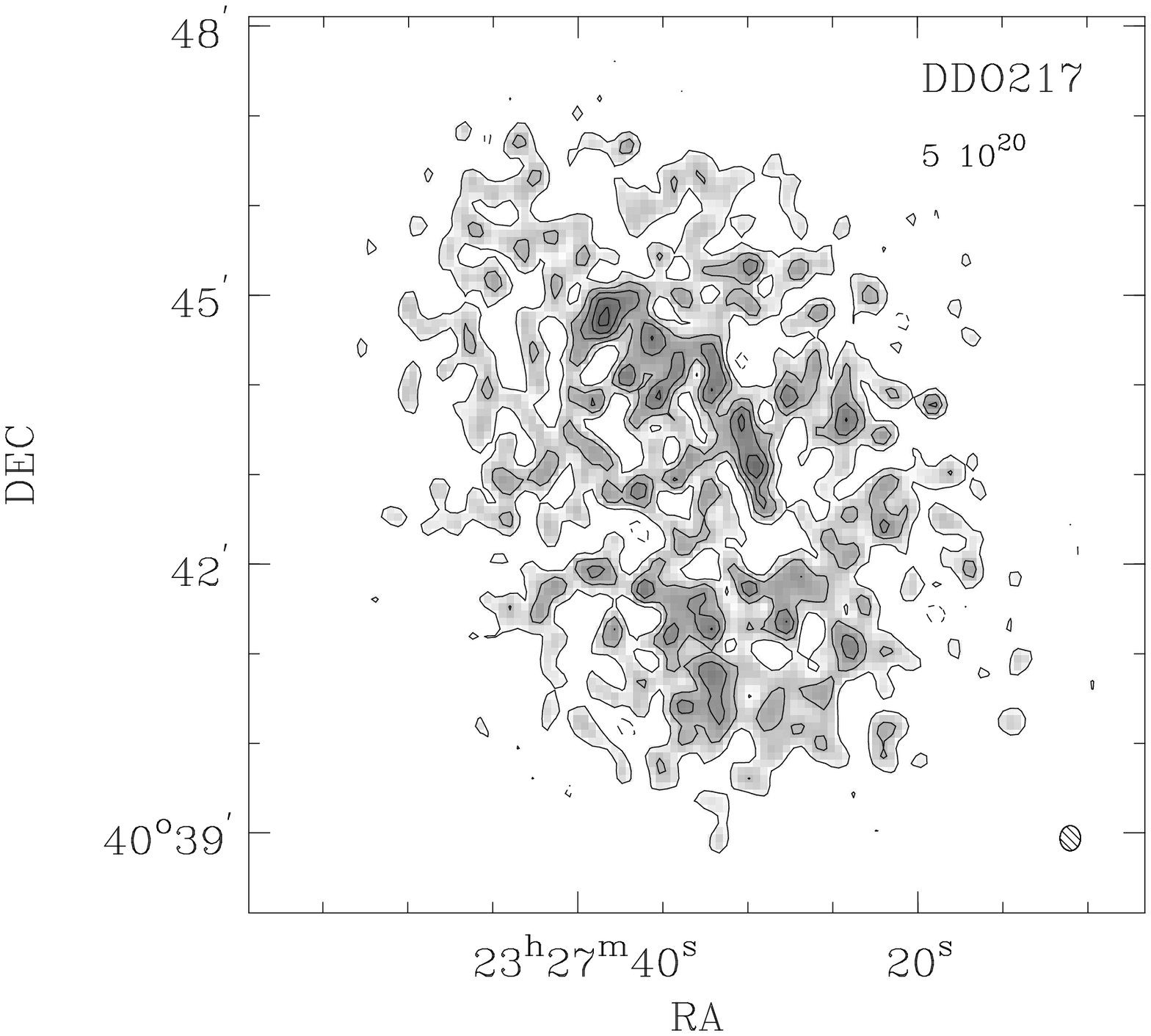}}
\end{minipage}
\hfill
\caption{continued}
\label{NHI-sample}
\end{figure*}

The line 
profiles resulting from this procedure are shown in 
Fig.~\ref{prof-sample} From these line profiles we calculated the HI 
flux integral $ FI=\sum_{v} S_v \Delta v_{c}$, the flux-density-weighted 
mean velocity $v_{\rm sys} = { {\sum_{v} v S_v } \over FI }$ and, 
assuming optically thin emission, the total HI mass $M_{HI} = 2.356 
\cdot 10^5 D^2 \sum_{v} S_v \Delta v_{c}$, where $D$ is the distance in 
Mpc, $S_v$ the flux density in Jy and $\Delta v_{c}$ the width of a 
single channel in $\kms$. $M_{\rm HI}$ is in units of solar mass. 
The resulting values are listed in Table~\ref{line-tab}. In that table,
we have also listed the ratios of the HI mass to the blue luminosity 
(defined as the ratio of the absolute blue magnitudes of the galaxy and
the Sun) and far-infrared luminosity (defined as $4 \pi D^{2} \times FIR$
respectively.

As we have already mentioned, the lack of the shortest spacings may cause 
us to underestimate the total flux contained in the interferometer map.
In order to verify the importance of this effect, we have compared
the WSRT flux-integrals just determined with flux-integrals based on
single dish measurements.  In Fig.~\ref{fluxflux} we have plotted the
values listed in the large compilation by Bottinelli et al. (1990)
as a function of the WSRT values. The errorbars do not take into account 
the (systematic) flux calibration uncertainty, which is about of 10$\%$. 
There is good agreement is good, but the WSRT fluxes tend to be lower
by factors up to 1.25. However, as an anonymous referee has pointed out, 
the values listed by Bottinelli et al. (1990) represent a very inhomogeneous 
sample, incorporating corrections for assumed HI extent and corrections
for telescope-deptendent HI flux scales. The magnitude of and uncertainty
in these corrections may easily be comparable to the difference in
integral values noted above. In an attempt to further investigate this
matter, we have also compared the WSRT integrals to those listed
by Tifft $\&$ Huchtmeier (1990) for seven of the eight galaxies
in common (excluding the relatively extended DDO~217). The single-dish
integrals listed by those authors were determined with the Effelsberg 
100 m and NRAO 300 ft telescopes (HPBW 9.3$'$ and 10$'$ respectively) on
practically identical flux scales; no correction for extent was applied.
We find that the seven WSRT integrals are on average 10$\%$ {\it higher} 
than the single-dish vales by Tifft $\&$ Huchtmeier 1990). This suggests
that at least these seven galaxies (a) are correctly represented by the 
WSRT maps, not suffering from missing flux problems and (b) should have 
overall HI sizes of order 3$'$ as indeed shown in Fig.~\ref{NHI-sample},
although the flux scale uncertainties are easily of the same magnitude.
 
The actual shape of an HI line profile depends on both the HI gas 
distribution and on the HI gas kinematics. For instance, a 
double-peaked line profile may result from a uniform surface density 
disk with a flat rotation curve, but it may also be the signature of 
a rotating ring. Inspection of Figs.~\ref{prof-sample} and 
~\ref{NHI-sample} shows that the HI distribution of galaxies having 
a double-peaked profile is indeed relatively homogeneous. In contrast, 
the line profiles of galaxies with a distinct central minimum ({\small 
DDO\,}63, {\small DDO\,}125, {\small DDO\,}165, Mk 178) are all 
single-peaked. Galaxies with a double peaked profile are on average more
luminous, but the low luminosity objects {\small DDO\,}52 and {\small
DDO\,}87 ($M_{\rm B} > -14$) also have a double-peaked line profile.
These objects have nearly flat rotation curves at the outermost
measured point, as we will show later.

\subsection{HI distribution in dwarf galaxies} 

We have produced maps of the HI distribution of the sample galaxies 
by summation of the individual cleaned channel maps. In order to minimize 
noise contributions of emission-free positions, we took from each 
channel map only the region contained within the cleaning mask. As a
consequence, the remaining noise is now a function of position in
the map. Assuming optically thin emission, we have converted HI line 
intensities into HI column densities (H atoms per $\cm2$), by using:
$$
N_{\rm HI}={ {1.104 \cdot 10^{21}} \over {b_\alpha b_\delta}} \int I_v dv 
$$
where $I_v$ is in mJy, $v$ is in $\kms$, and $b_\alpha$ and $b_\delta$ 
are the {\small FWHM} beamsizes (in arcsec) in right ascension and 
declination respectively.

In comparing HI column density maps, one should take into account
the linear resolution of the maps. For example, the small {\small DDO\,}43 
is, in fact, approximately the same linear size as the larger 
{\small DDO\,}47. The size of condensations at a particular column density
level relative to the overall HI distribution is a useful parameter in
comparisons of HI column density maps, because it is relatively
insensitive to the linear resolution. The greyscales in Fig.~4 are 
the same for all objects in order to facilitate such comparisons. 

The sample dwarf galaxies can be divided broadly into three classes based 
on the appearance of the high-column-density HI distribution.

A. Relatively small high-column-density regions are distributed
throughout the HI disk. Typical sizes are smaller than 100 pc. 
The average HI column density is low, and very high HI column density
regions ($\rm N_{\rm HI}>2 \cdot 10^{21}\ \cm2$) do not occur. 
On the whole, these objects are rather featureless.
The morphology resembles the HI distribution in Sc galaxies. 
{\small DDO\,}43, {\small DDO\,}47, {\small DDO\,}52, 
{\small DDO\,}73, {\small DDO\,}87, {\small DDO\,}123, {\small DDO\,}133 
and {\small DDO\,}217 all belong to this class.

B. A high-column-density region ($\rm N_{\rm HI}>10^{21}\ \cm2$) in the 
shape of a (sometimes broken) ring is present, comparable in size to the 
HI disk. The central parts of the galaxy have relatively low column
densities. Objects in this class are {\small DDO\,}46, {\small NGC\,}2537, 
{\small DDO\,}63, {\small NGC\,}2976, {\small DDO\,}68, {\small DDO\,}83, 
Mk~178, {\small DDO\,}125, {\small DDO\,}165 and {\small DDO\,}166.
 
C. The structure is dominated by a single prominent complex of very high 
column density ($\rm N_{\rm HI}>3 \cdot 10^{21}\ \cm2$). Galaxies in this 
class are {\small NGC\,}3738, Mk209, {\small DDO\,}168 and {\small DDO\,}190.

Some sample galaxies are left unassigned to a class, mainly because
they appear to be viewed edge-on: {\small DDO\,}22 (class B?), 
{\small DDO\,}48 (class A?), {\small UGC\,}4278, {\small DDO\,}64 
(class C?) and {\small DDO\,}185. {\small DDO\,}101 is too small. 
{\small DDO\,}216 is hard to assign, but may be closest to class B.

{\small DDO\,}47 is the only object in the sample with some evidence of 
spiral structure in the HI. Two arms appear on either side of the HI disk in
position angle $\pm 90^\circ$ (see also Puche \& Westpfahl 1993).

\end{document}